\documentclass[fleqn,usenatbib]{mnras}

\usepackage[T1]{fontenc}
\usepackage{ae,aecompl}

\usepackage{graphicx}	% Including figure files
\usepackage{amsmath}	% Advanced maths commands
\usepackage{amssymb}	% Extra maths symbols
\usepackage{pdflscape}
\usepackage{times,txfonts}
\usepackage{hyperref}	

\makeatletter
\renewcommand{\@printed}{}
\makeatother

\newcommand{\feh}{$\mathrm{[Fe/H]}$}
\newcommand{\afe}{$\mathrm{[\alpha/Fe]}$}
\def\red{}%\color{red}}

\title[Dynamical heating across the Milky Way disc]{Dynamical heating across the Milky Way disc using APOGEE and \emph{Gaia}}

\author[J. T. Mackereth et al.]{
J. Ted Mackereth,$^{1,2}$\thanks{E-mail: j.e.mackereth@bham.ac.uk (UoB)}
Jo Bovy,$^{3,4}$\thanks{Alfred P. Sloan Fellow}
Henry W. Leung,$^{3}$
Ricardo P. Schiavon,$^{1}$
\newauthor
Wilma H. Trick,$^{5}$
William J. Chaplin,$^{2}$
Katia Cunha,$^{6,7}$
Diane K. Feuillet,$^{8}$
\newauthor
Steven R. Majewski,$^{9}$
Marie Martig,$^{1}$
Andrea Miglio,$^{2}$
David Nidever,$^{10}$
\newauthor
Marc H. Pinsonneault,$^{11}$
Victor Silva Aguirre,$^{12}$
Jennifer Sobeck,$^{13}$
Jamie Tayar,$^{14}$\thanks{Hubble Fellow} 
\newauthor
and Gail Zasowski$^{15}$
\\
$^{1}$Astrophysics Research Institute, Liverpool John Moores University, 146 Brownlow Hill, Liverpool, L3 5RF, UK\\
$^{2}$School of Astronomy and Astrophysics, University of Birmingham, Edgbaston, Birmimgham, B15 2TT, UK\\
$^{3}$Department of Astronomy and Astrophysics, University of Toronto, 50 St. George Street, Toronto, ON M5S 3H4, Canada\\
$^{4}$Dunlap Institute for Astronomy and Astrophysics, University of Toronto, 50 St. George Street, Toronto, ON M5S 3H4, Canada\\
$^{5}$Max-Planck-Insitut f\"ur Astrophysik, Karl-Schwarzschild-Str. 1, D-85748 Garching b. M\"unchen, Germany\\
$^{6}$University of Arizona, Tucson, AZ 85719, USA\\
$^{7}$Observat\'orio Nacional, S\~ao Crist\'ov\~ao, Rio de Janeiro, Brazil\\
$^{8}$Max-Planck-Institut f\"ur Astronomie, K\"onigstuhl 17, D-69117 Heidelberg, Germany\\
$^{9}$Department of Astronomy, University of Virginia, Charlottesville, VA 22904, USA\\
$^{10}$National Optical Astronomy Observatories, Tucson, AZ 85719, USA\\
$^{11}$Department of Astronomy, The Ohio State University, Columbus, OH 43210, USA\\
$^{12}$Stellar Astrophysics Centre, Department of Physics and Astronomy, Aarhus University, Ny Munkegade 120, DK-8000 Aarhus C, Denmark\\
$^{13}$Department of Astronomy Box 351580, University of Washington, Seattle, WA 98195-1580, USA\\
$^{14}$Institute for Astronomy, University of Hawaii, 2680 Woodlawn Drive, Honolulu, Hawaii 96822, USA\\
$^{15}$Department of Physics \& Astronomy, University of Utah, Salt Lake City, UT, 84112, USA}
\date{Accepted XXX. Received YYY; in original form ZZZ}

\pubyear{2018}

\begin{document}
\label{firstpage}
\pagerange{\pageref{firstpage}--\pageref{lastpage}}
\maketitle

% Abstract of the paper
\begin{abstract}
The kinematics of the Milky Way disc as a function of age are well measured at the solar radius, but have not been studied over a wider range of Galactocentric radii. Here, we measure the kinematics of mono-age, mono-\feh{} populations in the low and high \afe{} discs between $4 \lesssim R \lesssim 13$ kpc and $|z| \lesssim 2$ kpc using 65,719 stars in common between APOGEE DR14 and \emph{Gaia} DR2 for which we estimate ages using a Bayesian neural network model trained on asteroseismic ages. We determine the vertical and radial velocity dispersions, finding that the low and high \afe{} discs display markedly different age--velocity-dispersion relations (AVRs) and shapes $\sigma_z/\sigma_R$. The high \afe{} disc has roughly flat AVRs and constant $\sigma_z/\sigma_R = 0.64\pm 0.04$, whereas the low \afe{} disc has large variations in this ratio which positively correlate with the mean orbital radius of the population at fixed age. The high \afe{} disc component's flat AVRs and constant $\sigma_z/\sigma_R$ clearly indicates an entirely different heating history. Outer disc populations also have flatter radial AVRs than those in the inner disc, likely due to the waning effect of spiral arms. Our detailed measurements of AVRs and $\sigma_z/\sigma_R$ across the disc indicate that low \afe{}, inner disc ($R \lesssim 10\,\mathrm{kpc}$) stellar populations are likely dynamically heated by both giant molecular clouds and spiral arms, while the observed trends for outer disc populations require a significant contribution from another heating mechanism such as satellite perturbations. We also find that outer disc populations have slightly positive mean vertical and radial velocities, likely because they are part of the warped disc.
%8996
\end{abstract}

% Select between one and six entries from the list of approved keywords.
% Don't make up new ones.

%%%%%%%%%%%%%%%%%%%%%%%%%%%%%%%%%%%%%%%%%%%%%%%%%%

%%%%%%%%%%%%%%%%% BODY OF PAPER %%%%%%%%%%%%%%%%%%

\section{Introduction}

%@arxiver{GaiaAPOGEE_agefehafe.pdf,sigma_z_R_age_rmean_alphasplit.pdf,vRz0_Rmean_age.pdf}
%\ted{Overall comment on the intro: a little long! Not a big deal, but if you could shorten that would be good.}

The present day kinematics of stars in the Milky Way are immutably tied to the dynamical history of the Galaxy. As a result, the measurement of the kinematics of the stellar disc as a function of age or element abundances offers tight constraints on models for the formation and evolution of the Galaxy \citep{2013A&ARv..21...61R}. In the new era of \emph{Gaia} \citep{2016A&A...595A...1G}, our access to the kinematics of stars has increased by orders of magnitude. In addition, the recent advent of large-scale spectroscopic surveys (e.g. Gaia-ESO, \citeauthor{2012Msngr.147...25G} \citeyear{2012Msngr.147...25G}; APOGEE, \citeauthor{2015arXiv150905420M} \citeyear{2015arXiv150905420M}; GALAH \citealt{2016arXiv160902822M}) has delivered stellar spectra, and therefore element abundance measurements, on an unprecedented scale. However, the measurement of accurate stellar ages, especially those of red-giant stars, has been problematic \citep[see][and references therein]{2010ARA&A..48..581S}. On this front, recent advances, combining spectroscopic surveys with high-quality asteroseismic data \citep[e.g.][]{2014ApJS..215...19P,2018arXiv180409983P}, have lead to great improvements in age measurements \citep[e.g.][]{2016MNRAS.456.3655M,2018MNRAS.475.5487S}, leading Galactic astrophysics into a truly multi-dimensional mode of operation.

Pioneering work which studied the variation of stellar velocities with indicators of their age demonstrated that there exists a positive correlation between age and velocity dispersion in the solar vicinity  \citep{1946ApJ...104...12S,1950AJ.....55..182R,1950ApJ...112..554R}. This property of the Galactic disc has since been well measured and characterised in the solar neighbourhood \citep[e.g.][]{1977A&A....60..263W,2007MNRAS.380.1348S,2008A&A...480...91S,2011A&A...530A.138C}. Observational limitations have meant that this relationship is not yet well constrained throughout the disc which would allow a detailed assessment of its dynamical history and the heating processes which have shaped it.  

 \citet{1967ApJ...150..461B} were among the first to show that transient spiral arms could contribute to disc heating, while \citet{1951ApJ...114..385S,1953ApJ...118..106S} demonstrated similar effects from small scale irregularities in the disc potential, such as those caused by Giant Molecular Clouds (GMCs). \citet{1990MNRAS.245..305J} considered the combined effect of both these heating agents, finding that increasingly prominent spiral perturbations (over those from GMCs) tend to flatten the axis ratio of velocity dispersions (e.g. decreasing $\sigma_{z}/\sigma_{R}$) over time. However, these models all struggled to reconcile the predicted slope of the vertical AVR \citep[usually modelled as a power law $\sigma_z \propto \tau^{\beta_z}$, where models predict $\beta_z \sim 0.25$, e.g.][]{1984MNRAS.208..687L,2002MNRAS.337..731H} with that observed at the solar vicinity \citep[where $\beta_{z}$ is closer to $\sim 0.5$, e.g.][]{1977A&A....60..263W,2007MNRAS.380.1348S,2008A&A...480...91S}. Increased heating from GMCs early in the history of the disc can be invoked to overcome these discrpeancies \citep[e.g.][]{2016MNRAS.462.1697A,2016MNRAS.459.3326A}, and a recent study of the vertical actions of APOGEE-\emph{Gaia} Red Clump giants as a function of age in the low \afe{} disc seemed to confirm this notion \citep{2018arXiv180803278T}.

High resolution cosmological zoom simulations have also revealed that other non-axisymmetric features such as bars may be a dominant heating agent, alongside (rare) perturbations from relatively massive ($M \gtrsim 10^{10}\ \mathrm{M_{\odot}}$) satellite mergers \citep{2016MNRAS.459..199G}. This is particularly prescient given that it has become apparent that the Milky Way likely underwent such a merger $\sim 10$ ago \citep[e.g.][]{2018MNRAS.478..611B,2018arXiv180510288D,2018arXiv180606038H,2018MNRAS.tmp.1537K,2018arXiv180800968M}. The effect of this accretion event on the dynamics of the disc itself is yet to be fully considered.

Understanding the effect of different heating agents is of course instructive, but any constraints therein on galaxy formation are limited without consideration of the velocity dispersion that the stars formed with. It is clear, from cosmological simulations at least, that discs do not tend to form thin and cool, but instead follow an "upside-down" formation \citep[e.g.][]{2004ApJ...612..894B,2012MNRAS.426..690B,2013MNRAS.428..129S,2013ApJ...773...43B,2013MNRAS.436..625S,2017MNRAS.467..179G,2016arXiv160804133M,2017arXiv170901040N}, where the stars form with the same velocity dispersion as the settling gas discs. Disentangling the combined outcome of the equilibrating gas disc and the subsequent heating of its stellar populations using observational data \citep[e.g.][]{2017MNRAS.472.1879L} will lead to the greatest insights into the evolution of the Milky Way. 

Whilst it has been informative to look at how the disc kinematics change with either age or abundances in isolation, it has become clear that stellar ages and element abundances in the solar neighbourhood (and beyond) do not simply correlate with one another, with perhaps the clearest example of this being \afe{}-\feh{} as a function of age \citep[e.g.][]{2016MNRAS.456.3655M,2016ApJ...823..114N}. The discovery of the separated high and low \afe{} disc components in the solar vicinity \citep[e.g.][]{1998A&A...338..161F,2000AJ....120.2513P}, the subsequent mapping of them through the galaxy \citep{2012A&A...545A..32A,2014A&A...564A.115A,2014ApJ...796...38N,2015ApJ...808..132H}, and the characterisation of their structure \citep{2012ApJ...751..131B,2012ApJ...753..148B,2013A&A...560A.109H,2016ApJ...823...30B}, has lead to a new view of our Galaxy. In particular, it seems that the commonly assumed linkage between the high and low \afe{} disc and the thick and thin disc is not as clear cut \citep{2012ApJ...751..131B}. In fact, any structural dichotomy between the high and low \afe{} discs is more apparent \emph{radially}, with the high \afe{} disc being centrally concentrated, and the low \afe{} populations occupying donut shaped annuli \citep{2016ApJ...823...30B}, which change shape as a function of age and metallicity \citep{2017arXiv170600018M}. \citet{2016A&A...589A..66H} present similar arguments, proposing an inner/outer disc divide to be more constraining to the evolution of the disc.

Models for the formation of the bimodality in \afe{} generally predict that high \afe{} stars form in rapid and intense star formation, precluding enrichment by Type Ia SNe. These high star formation efficiency environments are either brought about by rapid infall of gas \citep[e.g.][]{1997ApJ...477..765C,2001ApJ...554.1044C,2009IAUS..254..191C}, or are present in the innermost, high density regions of modelled discs \citep[e.g.][]{2009MNRAS.396..203S,2009MNRAS.399.1145S}. Alongside such analytic models, \afe{} bimodality has also been recently realised in fully self-consistent cosmological simulations. \citet{2017arXiv170807834G} found that \afe{} bimodality in the discs of the AURIGA zoom-in simulations \citep{2017MNRAS.467..179G} arises due to a double peaked star formation history, interjected by a shrinking of the gas disc. They also showed that \afe{} bimodality can appear without this process, via an initial rapid burst of star formation in the inner galaxy, but noted that the bimodality was not ubiquitous among the simulations. \citet{2018MNRAS.477.5072M} further showed that \afe{} bimodality in the EAGLE simulations \citep{2015MNRAS.446..521S,2015MNRAS.450.1937C} is very rare (occuring in $\sim 6\%$ of Milky Way mass galaxies), and driven largely by the atypical assembly history of the haloes hosting the bimodal galaxies, which accreted mass faster at earlier times (and slower at late times) than their non-bimodal counterparts. Importantly, they showed that low and high \afe{} populations form chemically separated from each other and usually overlap in age, becoming co-spatial later in the history of the galaxies. 

These models for the formation of the high and low \afe{} components make qualitative predictions for the kinematic structure of these populations, which are becoming more easily testable as the available data becomes more sophisticated. For example, the `radial migration' model of \citet{2009MNRAS.396..203S} predicts that outwardly migrating high \afe{} stars, formed in the inner disc at higher $\sigma_z$ {  act to increase the slope of the age-velocity dispersion relationship in the solar neighbourhood (relative to models where stars do not migrate)}. Therefore, isolating the low \afe{} disc AVR should yield a reconciliation with the relation predicted by models of time-dependent scatterers \citep[$\beta_z \sim 0.25$, e.g.][]{1984MNRAS.208..687L,2002MNRAS.337..731H}. `Two-infall' type models require a very rapid infall of gas onto the proto-Milky Way to form the high \afe{} stars, which may correspond to the clumpy galaxies observed at high redshift (e.g. \citeauthor{1998Natur.392..253N} \citeyear{1998Natur.392..253N}, and see also the recent preprint by \citeauthor{2019arXiv190100931C} \citeyear{2019arXiv190100931C}), and rapid merging at similar times seen in simulations \citep[e.g.][]{2004ApJ...612..894B,2009ApJ...707L...1B}, consistent with the Milky Way-like galaxies from \citet{2018MNRAS.477.5072M}. Rapid, early accretion and subsequent intense star formation would likely produce a centrally concentrated and very kinematically hot high \afe{} population. Such models require that the high and low \afe{} populations become co-spatial, and so require that the high \afe{} populations are heated radially somehow, bringing stars to the solar radius. The distinction in kinematics between radial migration and heating may be slight, but would place a very strong constraint on these models. At present, the main distinction between radial migration and heating can be made by measuring the flaring of the disc \citep[e.g.][]{2012ApJ...753..148B}, as it is expected that the flaring resultant from each process is different \citep[e.g.][]{2012A&A...548A.127M}. Understanding whether the flaring, the disc surface density, and its kinematics are commensurate with one another would help to disentangle these processes. %\ted{Should we mention flaring of the disk here? Expected for migration, maybe not for heating or at least differently?}

Since the second \emph{Gaia} data release \citep[DR2][]{2018arXiv180409365G}, there has been a rapid development in our understanding of the kinematics and dynamics of the disc. \citet{2018A&A...616A..11G} mapped the kinematics of the largest portion of the Milky Way to date, revealing a rich structure and unveiling the clear non-axisymmetries of the disc, providing the first insight into our new view of the Galactic disc. \citet{Antoja:2018aa} showed that \emph{Gaia} DR2 data reveals a `phase-spiral' feature in the solar neighbourhood, which is now widely considered to be the lasting harmonic relic of a recent perturbation by a satellite flyby \citep[e.g.][]{2018arXiv180902658B,2018MNRAS.481.1501B,2019MNRAS.tmp..574L}. Such a  wave-like oscillation is also apparent as an asymmetry in number counts as a function of height above the Galaxy midplane \citep{2018arXiv180903507B}. It is worth noting also that the effect of phase-mixing and warping on phase space in the disc was studied prior to \emph{Gaia} by a number of works \citep[e.g.][]{2009MNRAS.396L..56M,2009MNRAS.397.1599Q,2012MNRAS.419.2163G,2015MNRAS.454..933D}. The general picture which emerges is that the disc which is likely somewhat out of equilibrium due to recent events in its history.

As mentioned briefly above, a study of the vertical kinematics of Red Clump (RC) giant stars in common with APOGEE and \emph{Gaia} DR2 recently suggested that gradual orbit scattering by small perturbations like those from GMCs may be enough to explain the observed trends between vertical actions and age - showing that a model of birth temperature scaling with star formation rate did not fit the data well \citep{2018arXiv180803278T}. That study focused mainly on the younger, low \afe{} dominated populations, whereas we extend the modelling of the vertical kinematics to the high \afe{}, old stellar populations, as well as examining the radial kinematics. \citet{2018arXiv180803278T} found that the slope of vertical action-age relationship (which can be understood as the adiabatically invariant counterpart to the AVR) increased with Galactocentric radius, but found this increase to also be in-line with expectations from GMC scattering.

In this paper, we present a dissection of the disc kinematics in age, \feh{} and \afe{}, exploiting an unprecedented, multi-dimensional data set consisting of stars with element abundances and ages based on APOGEE spectra, whose distances are estimated to high precision using neural network modelling of these spectra trained on \emph{Gaia} DR2 data, which also offers high-quality proper motion information. In Section \ref{sec:data}, we present and describe this data set, and describe the procedure by which we estimate ages for a large portion of the APOGEE catalogue, using a Bayesian Convolutional Neural Network trained on APOGEE spectra and asteroseismic data from the APOKASC catalogue. We complete that section by describing the features seen in the data in the age-\afe{}-\feh{} plane. In Section \ref{sec:models}, we present the procedure used to model the velocity dispersions in the vertical and radial direction in the Galactic disc for mono-age, mono-\feh{} populations in the low and high \afe{} discs. Section \ref{sec:results} presents the main results from modelling the velocity dispersion of the mono-age, mono-\feh{} populations. In Section \ref{sec:discussion}, we assess the results in the context of previous work on constraining the effects of disc heating, and discuss the implications of our findings on models for the origin of the high and low \afe{} populations. We summarise and conclude the paper in Section \ref{sec:conclusion}.

% \begin{itemize}
% \item The recent second data release from the \emph{Gaia} satellite has unlocked access to velocity information for stars on an unprecedented scale... [RECENT RESULTS]
% \item It is well established that the nearby disc stars of the Milky Way divide into two groups in \afe{}-\feh{} space... [BIMODALITY RESULTS, CONNECTION TO KINEMATICS]
% \item Mapping the velocity structure of the disc has already proven a fruitful method of discerning aspects of the history and nature of the Milky Way disc... [TALK ABOUT AVR RESULTS, GMC HEATING ETC]
% \item Many of the above studies have focused on the more recent heating history of the disc, and do not study the interface between the young and old disc populations... [SIMULATION RESULTS?]
% \item In this paper, we....
%\end{itemize}

\section{Data}
\label{sec:data}
We use a catalogue of stellar positions, velocities, element abundances and estimated ages, comprising a cross match of the catalogues from the fourteenth data release \citep[DR14,][]{2018ApJS..235...42A} of the SDSS-IV APOGEE-2 survey, and the second data release \citep[DR2,][]{2018arXiv180409365G} of the ESA-\emph{Gaia} mission. Ages are estimated from a neural network based model (described below) trained on data from the APOKASC catalogue, which contains stars observed both spectroscopically by APOGEE and asteroseismically by the \emph{Kepler} mission. 

\subsection{The APOGEE-2 DR14 catalogue}

\label{sec:APOGEE}

APOGEE \citep{2015arXiv150905420M} is a spectroscopic survey of the Milky Way in the near infra-red H-band (1.5-1.7$\mathrm{\mu}$m), which has observed over 200,000 stars at high signal-to-noise ratio (SNR > 100 pixel$^{-1}$) and high resolution (R $\sim$ 22 500), measuring over 15 element abundances. We use the DR14 data \citep{2018ApJS..235...42A}, which consists of a combination of stars observed between the first and second iterations of the survey, APOGEE-1 \citep[forming part of SDSS-III; ][]{2011AJ....142...72E} and 2 \citep[in SDSS-IV; ][]{2017AJ....154...28B}. In this paper we refer to these surveys collectively as APOGEE. All APOGEE data products used in this paper are those output by the standard data analysis pipeline, the APOGEE Stellar Parameters and Chemical Abundances Pipeline \citep[ASPCAP][]{2016AJ....151..144G}, which uses a precomputed spectral library \citep{2015AJ....149..181Z}, synthesised using a customised $H$-band linelist \citep{2015ApJS..221...24S}, to measure stellar parameters and element abundances. A full description and examination of the analysis pipeline is given in \citet{holtzdr14}. The individual element abundances are well tested against samples from the literature \citep{jonssondr14}, and found to agree very well.

We refer here only to the abundances of the $\alpha$-elements included in DR14: Oxygen, Magnesium, Silicon, Sulphur, and Calcium, and the abundance of Iron. All abundances are those calculated by ASPCAP, and included in the APOGEE catalogue. We combine the $\alpha$ elements to attain the ratio of the mean $\alpha$ element abundance to that of Iron, \afe{}. The element abundances are determined by ASPCAP via a two-step process. The best fit stellar parameters $T_{\mathrm{eff}}$, $\log(g)$, $\nu_{\mu}$ (microturbulent velocity), [M/H], $\mathrm{[\alpha/M]}$, [C/M] and [N/M] are determined by a global fit to the grid of synthetic stellar spectra. Individual element abundances are then estimated by fitting windows in the spectrum to synthetic spectra with varying [M/H] (or $\mathrm{[\alpha/M]}$, [C/M] or [N/M], for $\alpha$-elements, Carbon and Nitrogen, respectively). For DR14, a small external calibration is applied to all the abundances, which forces the abundance ratios of solar metallicity stars in the solar vicinity to be equal to solar \citep{holtzdr14}.

We do not use Gaia parallaxes directly as distance indicators, but instead use spectro-photometric distances obtained by training a neural network to predict the luminosity of a star from its infrared spectrum using luminosities for a training set obtained from \emph{Gaia} DR2. This procedure is described in detail in \citet{2019arXiv190208634L}, but we re-iterate it here for clarity. The neural network has a similar architecture to the \verb|ApogeeBCNN()| network used in \citet{2018arXiv180804428L}: it is composed of 2 convolutional layers and 2 dense layers with rectifier activation (which introduce non-linearity in the hidden layers and prevent non-physical negative luminosity in the output); the network is optimized with the ADAM optimizer \citep{ADAM} and is implemented in \texttt{astroNN} \citep{2018arXiv180804428L}, which itself relies on the \texttt{TensorFlow} framework \citep{tensorflow}. To train the network in a way that allows us to take the \emph{Gaia} parallax uncertainty into account, {  we follow \citet{2018AJ....156..145A} and define the output to be the following transformation of the absolute magnitude $M_{K_s}$}

\begin{align}
	L_{\mathrm{fakemag}} & = 10^{\frac{1}{5}\,M_{K_s}+2}\,,\\
	& = \varpi 10^{\frac{1}{5}\,K_{s,0}}\,,\label{eq:fakemag2}
\end{align}

where $\varpi$ is the parallax in units of mas and $K_{s,0}$ is the extinction-corrected apparent magnitude in the $K_s$ band (for all stars, we obtain extinction corrections directly from the APOGEE catalogue). We then train the neural network to predict the {  'pseudo-luminosity'} $L_\mathrm{fakemag}$ from the continuum-normalized APOGEE spectrum using a $\chi^2$ objective function that takes the uncertainty in the $L_\mathrm{fakemag}$ of the training set (due to the uncertainty in $\varpi$), into account. When training the neural network, we adopt a $0.0562$ mas \emph{Gaia} DR2 parallax offset, because this value leads to approximately unbiased luminosities at small $\varpi$ \citep[see][for more details on this procedure]{2019arXiv190208634L}. We train the network using APOGEE DR14 \citep{holtzdr14} spectra that have SNR$>200$, no \texttt{APOGEE\_STARFLAG} flags set, radial velocity scatter smaller than $1\,\mathrm{km\,s}^{-1}$, and $\varpi_{Gaia}/\sigma_{\varpi,Gaia}>1$.

Once trained, we use the network to predict luminosities $L_\mathrm{fakemag}$ for the entire APOGEE DR14 sample and, combined with extinction-corrected $K_{s,0}$, we obtain spectro-photometric parallaxes using Equation \eqref{eq:fakemag2}. We obtain uncertainties on the predicted $L_\mathrm{fakemag}$ using dropout variational inference as an approximation of a Bayesian neural network \citep{2018arXiv180804428L} and these uncertainties are propagated to the spectro-photometric parallax. {  The distances are well tested in \citet{2019arXiv190208634L} against the APOGEE red clump star (RC) catalogue \citep{2014ApJ...790..127B}, alternative spectrophotometric distances from \citet{2016A&A...585A..42S} and also those from \citep{2018arXiv181009468H}, and found to be in very good agreement, if not with improved precision. We also test the spectrophotometric distances against the recently published catalogue of distances of \citet{2019arXiv190202355S}. We find that the distances agree to within $\sim 10\%$, comparable with the uncertainties quoted in the \citet{2019arXiv190208634L} catalogue.}

\subsection{\emph{Gaia} DR2}

\emph{Gaia} DR2 contains astrometric parameters for over 1.3 billion sources, collected over 22 months from July 2014. Many major improvements were made over the initial data processing, released in the DR1 and Tycho-Gaia Astrometric Solution (TGAS) catalogues, which are summarised in \citet{2018arXiv180409365G}. In particular, \emph{Gaia} DR2 is the first data release which is not tied to any external catalogue to provide proper motion measurements (DR1 was tied to the \textsc{Hipparcos} and \emph{Tycho}-2 catalogues), and uses its own refernce frame based on quasars \citep[\emph{Gaia}-CRF-2, ][]{2018A&A...616A..14G}. Other improvements included better modelling of the spacecraft attitude, and a significant improvement to the source detection algorithm which limits the number of spurious sources, and removes many duplicate sources which were included in the DR1 source catalogue (for this reason the catalogues between data releases are treated separately). As well as removing spurious sources, these improvements have led to significant increases in the astrometric precision, affording {  typical residuals on the astrometric solution} of $\sim 0.2$ to 0.3 mas in the middle of the magnitude range, decreasing to $\sim 2$ mas for the faintest sources.

We cross match the full APOGEE DR14 catalogue with the \emph{Gaia} DR2 source catalogue, searching in a cone of {\red radius} $0.5"$ around each source for its corresponding object in the $\emph{Gaia}$ catalogue using the CDS X-match service\footnote{\url{http://cdsxmatch.u- strasbg.fr/xmatch}}. We find that there are 254,789 objects in common between APOGEE and \emph{Gaia} DR2. After removing stars from APOGEE which have warning or bad flags set, selecting only stars which have proper-motion measurements from \emph{Gaia}, and removing duplicate entries from the APOGEE catalogue, 83,189 stars with 6D phase space information remain. In the following analysis, we only use stars with $1.8 < \log(g) < 3.0$ (removing dwarf and subgiant stars, and upper RGB stars) and those with \feh{} $> -0.5$, below which the APOKASC catalogue (used in the following section for training our age estimation model) has poor coverage. In addition to this, it was recently shown that extra mixing becomes significant at approximately this metallicity, which would substantially change the relationship between mass (and therefore the inferred ages) and $\mathrm{[C/N]}$ (Shetrone et al., in prep.), which is likely to be the parameter which drives the relationship between the spectroscopic data and age. This final sample contains 65,719 red giant stars, whose spectra are measured with a minimum SNR of $\sim 80$, and a median SNR of $\sim 170$. The median uncertainty on the proper motion measurements from \emph{Gaia} is $\sim 0.5$ mas, corresponding to a median uncertainty (after transformation) on the Galactocentric radial and vertical velocities of $\sim 2$ and $\sim 1\ \mathrm{km\ s^{-1}}$, respectively. The median uncertainty on the distances from \texttt{astroNN} (described in Section \ref{sec:APOGEE}) is also remarkably low, at $\sim 0.24$ kpc, better than 10\% in most cases.

\begin{figure}
\includegraphics[width=\columnwidth]{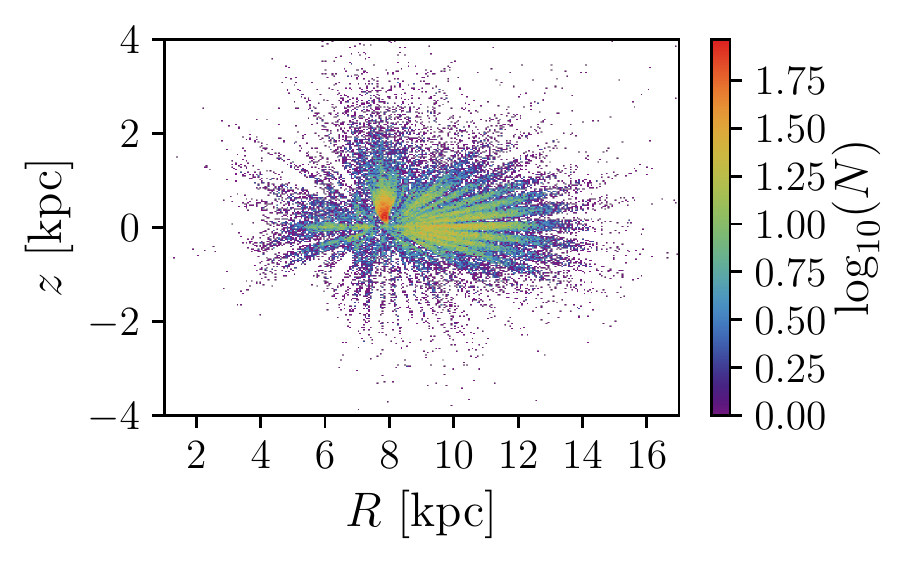}
\caption{\label{fig:rz} The distribution of the APOGEE-\emph{Gaia} sample used in this paper in Galactocentric $R$ and $z$. The pencil-beam selection of APOGEE is clearly apparent. The data cover a region of the Galaxy between $4\lesssim R \lesssim 13$ kpc and $-1 \lesssim z \lesssim 2$ kpc.}
\end{figure}

We transform the positions, proper motions, and radial velocities into the Galactocentric cylindrical frame, adopting the radial and vertical solar motion relative to the local standard of rest of \citet{2010MNRAS.403.1829S} and the tangential motion of the sun relative to the Galactic center of $245.6\ \mathrm{km\ s^{-1}}$ computed using the proper motion and distance to Sgr A* \citep{2004ApJ...616..872R,2018A&A...615L..15G}.  %and assuming a circular velocity at the solar radius $v_{\mathrm{c,0}} = 220\ \mathrm{km\ s^{-1}}$ \citep{2012ApJ...759..131B}
We propagate the observational uncertainties and their covariance matrix into this frame also. The data extends in Galactocentric radius roughly between $4 \lesssim R \lesssim 13$ kpc, and between $-1 \lesssim z \lesssim 2$ kpc above and below the plane (here, and throughout the paper, we assume $R_0 = 8$ kpc and $z_0 = 0.025$ kpc). We show the distribution of the data in $R$ and $z$ in Figure \ref{fig:rz}. This figure demonstrates the strong spatial selection biases which are present when using APOGEE data. For example, there is a large overdensity of stars above the plane, just inside the solar radius, that corresponds to the heavily-observed \emph{Kepler} field. Correcting for these biases is possible, but not necessary for the kinematic modelling performed here, because neither the APOGEE nor \emph{Gaia} data are kinematically biased. As we are interested only in modelling the velocity distributions as a function of position, these spatial biases do not affect our procedure.

\subsection{Asteroseismic data and age estimates for APOGEE red giants}

To make estimates of the ages of stars in the APOGEE-\emph{Gaia} data described above, we make use of data from the APOKASC-2 catalogue \citep{2018arXiv180409983P} of asteroseismic data for stars in common between APOGEE and \emph{Kepler} \citep{2010Sci...327..977B}. We use the APOKASC-2 data set to train a Bayesian Convolutional Neural Network (BCNN) model, as implemented in the \texttt{astroNN} python package \citep{2018arXiv180804428L}, to predict stellar ages from the APOGEE spectra. Our technique for modelling ages using the BCNN is described fully in Appendix \ref{sec:appA}. Briefly, we expect that the ability of the BCNN model to predict ages from the spectra to a good degree of accuracy stems from the presence of molecular bands of Carbon and Nitrogen in the APOGEE spectra. It is well documented in the literature that there is likely to be a relationship between these element abundances and the stellar mass \citep[and therefore age, e.g.][]{2015A&A...583A..87S}, and this has been exploited in previous work to estimate ages from APOGEE abundances and spectra \citep{2016MNRAS.456.3655M,2016ApJ...823..114N,2019MNRAS.484..294D}. The BCNN uses the full information content of the APOGEE spectra to estimate the ages, and allows for the proper propagation of uncertainties into the analysis, to predict ages with good accuracy and with reasonable error estimation. Using the method, we generate a catalogue of ages with a median uncertainty between $\sim 30$ and $35\%$ accross the full range of ages. We fully discuss the limitations of the predicted ages in Appendix \ref{sec:appA}, but note here for clarity that we expect that ages above $\sim 10$ Gyr are likely to be underestimated, and subject to very large errors, such that it becomes difficult to distinguish, for example, an 11 Gyr old star from one at 13 Gyr old. 

 \begin{figure}
\includegraphics[width=\columnwidth]{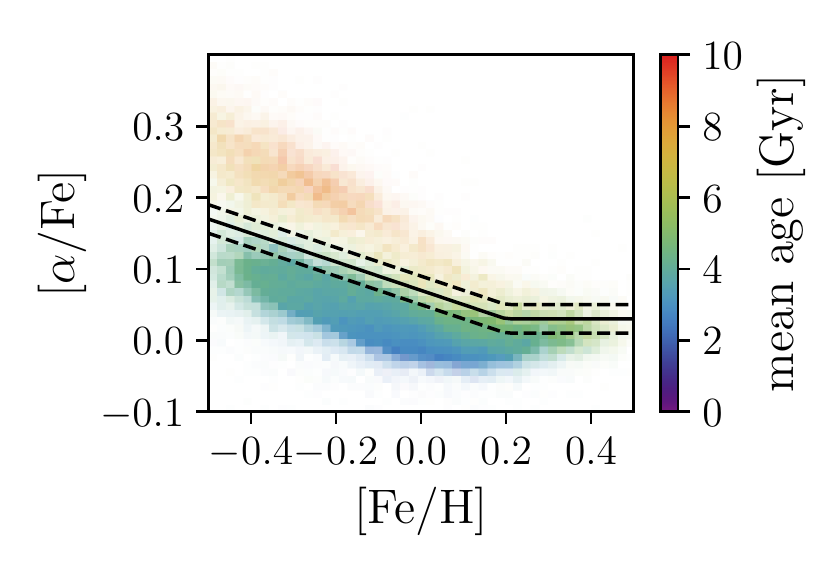}
\caption{\label{fig:afefehage} A 2D histogram of \feh{}-\afe{} for APOGEE DR14, coloured by the mean age in each bin, as modelled using the Bayesian Convolutional Neural Network implemented in \texttt{astroNN}, trained on APOKASC DR2 data. The transparency of the bins reflects the relative density of stars in that region of \afe{}-\feh{} space. The solid and dashed lines demonstrate the delineation we adopt in this paper to separate the high and low \afe{} populations in the disc. The stars inside the dashed region are excluded from the analysis, to avoid contamination between the two populations. The high \afe{} population is generally old, whereas the low \afe{} population has a spread in age.}
\end{figure}

\subsubsection{The age-\afe{}-\feh{} relation of disc stars}

\begin{figure}
\includegraphics[width=\columnwidth]{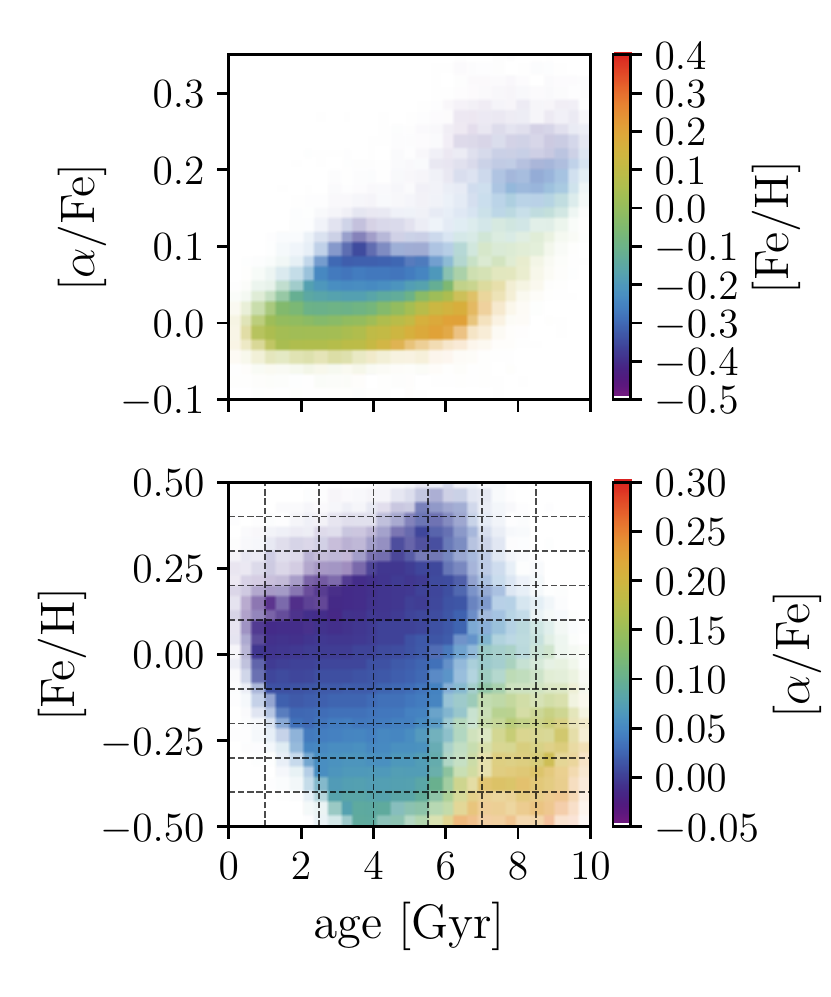}
\caption{\label{fig:agefehafe} The age-\afe{} (\emph{top}) and age-\feh{} (\emph{bottom}) relations for disc stars in APOGEE DR14, represented as two-dimensional histograms. In each panel, the bins are coloured by the abundance on the y-axis in the other panel, illustrating the complex, multi-dimensional nature of the data. The transparency of the bins reflects the relative density of stars in that part of the parameter space. The binning in age-\feh{} space which is adopted to perform the velocity modelling is demonstrated by the dashed grid. }
\end{figure}

For completeness, in Figures \ref{fig:afefehage} and \ref{fig:agefehafe} we demonstrate the complexity of the sample in age, \afe{} and \feh{} space. Figure \ref{fig:afefehage} shows the \afe{}-\feh{} plane as a 2D histogram, coloured by the mean age of the stars in each bin. The {  opacity} of the bins reflects the density of stars in that region of \afe{}-\feh{} space. The solid and dashed lines indicate the `by-eye' cut used to define high and low \afe{} stars later in the paper. Stars inside the dashed lines, the gap between which is equal to the median uncertainty in \afe{} ($\sim 0.05$ dex), are not used in the analysis, to avoid contamination of either sample by the other. The high \afe{} population is dominated by stars $\gtrsim 7$ Gyr old, whereas the low \afe{} population has a wide range of ages from the youngest at $\sim 1-2$ Gyr up to $\sim 6-7$ Gyr.

The age-\afe{} relation, coloured by \feh{}, is shown in the top panel of Figure \ref{fig:agefehafe}. The age-\feh{} relation, coloured by \afe{}, is shown in the lower panel. There are a number of features evident in these planes which warrant discussion, and provide a useful context for the further results which we present in the paper. The age-\afe{} relationship \emph{at the solar radius} has been shown to have very little scatter at old ages, extending to high \afe{} in a tight sequence, which becomes slightly more scattered at low \afe{} and young age \citep[e.g.][]{2013A&A...560A.109H}. {\red However, this appears to be dependent on the sample and age measurement method used, as asteroseismic analysis of stars in the \emph{Kepler} field (within 2 kpc of the Sun) suggests the age-\afe{} relationship may actually be broader at high \afe{} \citep{2018MNRAS.475.5487S}}. Our extension of the data to a much wider range of Galactocentric radii somewhat complicates this picture {\red and provides some insight into this (as yet, unsettled) issue}. The scatter in \afe{} is increased at all ages, and a number of interesting features become apparent. 

First, it is clear from the top panel that the high \afe{} `sequence', as is well characterised in the literature \citep[e.g.][]{2005A&A...433..185B,2015ApJ...808..132H,2014ApJ...796...38N}, which is apparent here extending between $6 \lesssim \mathrm{age} \lesssim 10$ Gyr, \emph{does} overlap slightly in age with the low \afe{} stars with ages $\sim 6$ Gyr old. The spread in \afe{} of old stars ($\gtrsim 6$ Gyr) is significant, and bridges the gap between the low and high \afe{} stars, with intermediate stars assuming roughly solar values of \feh{}. This high \afe{} sequence feature manifests itself in the lower panel as the old `branch' that extends from $\mathrm{(age, [Fe/H])}\sim(8,-0.4)$ to $\mathrm{(age, [Fe/H])}\sim(7,0.0)$, joining the large `cloud' of low \afe{} stars at the most metal-rich end (the low \afe{}, \feh{} rich, old blob at the lower-right of the low \afe{} population in the top panel). {  We note here that the spread in age of the oldest high \afe{} stars is likely to be inflated by the age uncertainties and therefore, any trends discussed among these populations is subject to this significant caveat.} {\red Our results do appear, however, to be qualitatively similar to those of \citet{2018MNRAS.475.5487S}, and show a spread in age at fixed \afe{} for higher \afe{} stars. }

The low \afe{} stars which are identifiable separately from the high \afe{} `sequence' in the top panel demonstrate a number of interesting features in both projections. In age-\afe{}, it is clear that there is some scatter in \afe{} at fixed age, most evident at $\mathrm{age} \sim 4$ Gyr, which is significant over the characteristic uncertainties on the \afe{} abundances ($\sim 0.05$ dex). This spread correlates with \feh{}, such that the stars with slightly enhanced \afe{} have the lowest values of \feh{}. These \afe{} enhanced low \afe{} population stars are evident in the lower panel at $\mathrm{(age, [Fe/H])}\sim(4,-0.4)$, with a similar \feh{} as the oldest stars in the high \afe{} population. The oldest stars in the low \afe{} population exhibit the highest \feh{} values. These old, low \afe{} population stars are clear in the lower panel at $\mathrm{(age, [Fe/H])}\sim(5,0.4)$. Understanding the nature of these stellar populations is likely to shed some light on the history of the disc.

In the following analysis, we aim to put these features into the context of the evolution of the Milky Way disc, using their kinematics to inform their origins and subsequent evolution. We will show that the different populations present in these planes are the result of a complex history of disc formation, which we can begin to disentangle and re-assess using the multi-dimensional data set at hand. {\red A future paper will complement the work here by studying the detailed orbital properties of these stars}.

\section{Modelling the Galactic kinematics of mono-age, mono-\feh{} populations}
\label{sec:models}
We model the radial and vertical velocity dispersion profiles of mono-age, mono-\feh{} populations as multivariate Gaussian distributions, explicitly fitting for the covariance terms, allowing us not only to assess the variation of the velocity dispersion as a function of age and \feh{}, but also to study variations in the tilt of the velocity ellipsoid with these parameters. This methodology also allows us to account for outliers and readily convolve the model with the observational uncertainties which, for \emph{Gaia}, can be strongly correlated with one another. 

We model the radial and vertical velocity dispersion with smoothly varying functions of $R$ and $z$, allowing them to vary exponentially in the $R$ direction, and following a quadratic relation in the $z$ direction (as in \citealt{2012ApJ...755..115B}) such that
\begin{equation}
\begin{aligned}
\sigma_{[R,z]}(R,z) & = \sigma_{[R,z]}(z,R_0 | a_{[R,z]}, b_{[R,z]}, \sigma_{[R,z]}(R_0, z_{1/2})) \\ & \times \exp{\left(\frac{-(R-R_0)}{h_\mathrm{\sigma_{[R,z]}}}\right)} + \delta^2_{v_{[R,z]}, i}
\end{aligned}
\end{equation}
{  where $a_{[R,z]}$ and $b_{[R,z]}$ are, respectively, linear and quadratic coefficients in the quadratic function
\begin{equation}
\begin{aligned}
    \sigma_{[R,z]}(z,R_0) = a_\mathrm{[R,z]}(z-z_{1/2})^2 + b_\mathrm{[R,z]}(z-z_{1/2})+\sigma_{\mathrm{[R,z]}}(R_0,z_{1/2})
\end{aligned}
\end{equation}which is centered on the median $z$ value, $z_{1/2}$.} Therefore, $\sigma_{[R,z]}(R_0,z_{1/2})$ gives the velocity dispersion at $z_{1/2}$ and $R_0$ (which here we define as the solar radius, assumed to be 8 kpc). The $R$ dependence is presumed to be exponential, with a scale length $h_{\sigma_{[R,z]}}$. The addition of $\delta^2_{v_{[R,z]}, i}$ accounts for the convolution of the model with the error on the velocity measurements. These assumptions on the $R$ and $z$ dependence of the velocity dispersion are not intended to be prescriptive of the actual underlying dependence, for which we have little prior expectation, but are intended to approximate it with smooth functions.

We assume that the tilt angle $\alpha$ is a function of $R$ and $z$ such that
\begin{equation}
\tan{(\alpha)} = \alpha_0 + \alpha_1 \frac{z}{R}\,,
\end{equation}
which includes the cases of a velocity ellipsoid that is always aligned with the Galactocentric cylindrical coordinate system ($\alpha_0 = \alpha_1 = 0$) or the Galactocentric spherical coordinate system ($\alpha_0 = 0$ and $\alpha_1 = 1$). The covariance between the radial and velocity dispersion is then
\begin{equation}
\sigma^{2}_{Rz}(R,z) = \left(\sigma^{2}_{R}(R,z)-\sigma^{2}_{z}(R,z)\right)\frac{\tan\alpha}{(1-\tan^2\alpha)}+\delta^2_{v_{Rz}, i}
\end{equation}
where we again convolve the model with the covariance between the radial and vertical velocity $\delta^2_{v_{R,z}, i}$. This then allows the construction of the covariance matrix which describes the velocity ellipsoid in the $R$ and $z$ directions
\begin{equation}
\Delta = \begin{bmatrix} \sigma^{2}_{R}(R,z) & \sigma^{2}_{Rz}(R,z) \\
				       \sigma^{2}_{Rz}(R,z) & \sigma^{2}_{z}(R,z) \end{bmatrix}.
\end{equation}
The likelihood of the parameters $O=[\sigma_{[R,z]}(R_0,z_{1/2}), a_{[R,z]}, b_{[R,z]}, h_{\sigma_{[R,z]}}, \alpha_0, \alpha_1, v_{R,0}, v_{z,0}, \epsilon]$ given the data for a given age-\feh{} bin can then be expressed as %\ted{First term should have $1/(2\pi)$ as well to be properly normalized, but is this in the model? Is it in $p_\mathrm{backgr.}$?}
\begin{equation}
\ln \mathcal{L}(O|R,z,v_R,v_z) = \sum_i{\ln\left[\frac{(1-\epsilon)}{2\pi|\Delta|^{1/2}}\exp{\left(-\frac{1}{2}\vec{v}_i\Delta^{-1}\vec{v}_i\right)}+\epsilon p_\mathrm{backgr.}(z,R)\right]}
\end{equation}
where $p_{\mathrm{backgr.}}(R,z)$ describes the velocity distribution of an interloper model that is contributed by a fraction of stars $\epsilon$. We use an interloper model which is also a normalised multivariate Gaussian, but with no covariance terms and with $\sigma_R = \sigma_z =100 \,\mathrm{km\ s^{-1}}$. This interloper model is also convolved with the observational uncertainty for each star. We find that for all mono-age, mono-\feh{} bins, $\epsilon$ is less than a few percent. {  The velocity distribution is a multivariate Gaussian centered on $v_{[R,z],0}$, where  $v_{[R,z],0}$ are allowed to vary freely. We find that allowing this variation in $v_{[R,z],0}$ provides a better fit to the data, and also reveals information regarding the non-axisymmetric velocity structure of the disc.}

We minimise the negative log-likelihood using a downhill
simplex algorithm \citep{doi:10.1093/comjnl/7.4.308}, and use this
optimal solution to initiate an Markov Chain Monte Carlo (MCMC)
sampling of the posterior PDF of the parameters O using an
affine-invariant ensemble MCMC sampler \citep{goodmanweare2010} as
implemented in the python package \texttt{emcee}
\citep{2013PASP..125..306F}. In the following, all reported parameter values and uncertainties are the median and standard deviation of the resulting MCMC chain. For all MCMC fits, we use 100 `walkers' over 500 iterations, generating 50,000 samples, of which we remove 300 from each chain, leaving 20,000 clean samples of the posterior PDF in each age-\feh{} bin. Cutting many samples allows the MCMC chains to be very well `burnt-in', ensuring an accurate sampling of the underlying PDF. 

We perform this fitting procedure across 8 bins in age with width $1.5$ Gyr between $1 < \mathrm{age} < 13$ Gyr, and 10 bins in \feh{} of width 0.1 dex, between $-0.5 < \mathrm{[Fe/H]} < 0.5$ dex (the adopted age-\feh{} bins are shown in Figure \ref{fig:agefehafe}), for stars in the low and high \afe{} populations, as defined in Figure \ref{fig:afefehage}. Modelling is only performed for bins with $N > 200$ stars to ensure good sampling of the underlying velocity distribution. This results in 35 populations modelled in the low \afe{} population and 13 in the high \afe{}. In both the low and high \afe{} populations, the oldest bin ($11.5 < \mathrm{age} < 13$ Gyr) is never sufficiently filled at any \feh{} and so we disregard it in the further discussion, reminding the reader that it is likely that the ages of the oldest stars in our sample ($\sim 10-11$ Gyr) are likely underestimated.{  The median number of stars in each bin for the low \afe{} populations is 1078, with a minimum of 204, and a maximum of 3916. The median number of stars in high \afe{} bins is lower, at 503, with a maximum of 771 and a minimum of 227. We test the robustness of the fits inside the low N bins by fitting stars in a test age-\feh{} bin (in the low \afe{} population, at $2.5 < \mathrm{age} < 4.0\ \mathrm{Gyr}$ and $-0.2 < \mathrm{[Fe/H]} < -0.1$ dex), using 1500,1000,500 and 200 random samples of stars from the bin to perform the fit. We find that the best fit values of all of the parameters do not vary, within their uncertainties, due to the lower number statistics. Thus, we can confidently use the bins with $N > 200$ stars. }

{  We demonstrate the resulting posterior PDFs for parameters $O$ for an example bin in age-\feh{} in Appendix \ref{sec:appB}.}
\section{Results}
\label{sec:results}
\subsection{Velocity dispersion profiles}

\label{sec:dispprofile}
\begin{figure*}
\includegraphics[width=\textwidth]{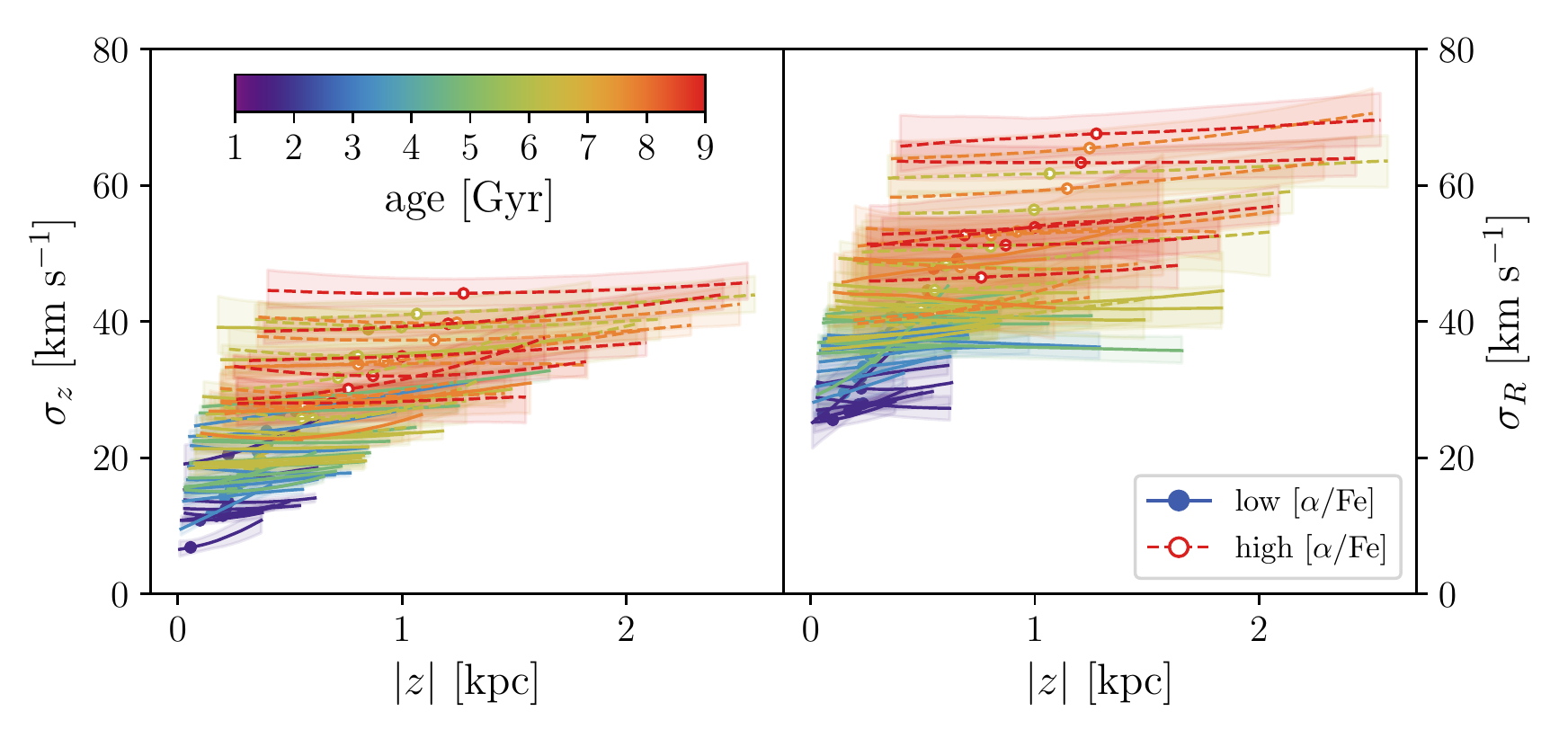}
\caption{\label{fig:sigfzage}The vertical (\emph{left}) and radial (\emph{right}) velocity dispersion of mono-age, mono-\feh{} populations fit to the {\red low (\emph{solid lines}) and high (\emph{dashed lines})} \afe{} sub-samples as a function of height above the midplane. The colour of each profile gives the age of the population, for which the profiles are only shown at $|z|$ between where 25 and 75 percent of the observed population are. The median $|z|$ of each population is indicated by the point on each line, and the coloured bands show the $1\sigma$ uncertainties at each $|z|$. In general, the profiles are flat as a function of $|z|$, indicating that mono-age populations are roughly isothermal, except for the youngest, coldest populations. The radial and vertical velocity dispersion increases as a function of age.}
\end{figure*}

We first examine the velocity dispersion profiles for mono-age mono-\feh{} populations. The radial and vertical velocity dispersion profiles as a function of height above the midplane $|z|$ at $R_0$ for each mono-age mono-\feh{} population are shown in Figure \ref{fig:sigfzage}. We show the variation of $\sigma_z$ (\emph{left}) and $\sigma_R$ (\emph{right}) for each mono-age mono-\feh{} population between the 5th and 95th percentile of the $|z|$ distances for that bin. The median height $z_{1/2}$ is indicated for each population by the scatter points. Although these profiles were not constrained to be flat, and $\sigma_R$ and $\sigma_z$ could vary smoothly with $|z|$, we find that a generally flat profile is fit in almost all cases, within the uncertainties, {  meaning that for most bins, the parameters $a_{[R,z]}$ and $b_{[R,z]}$ are fit to be very small, and with low uncertainty}. This means that these populations are well approximated as isothermal discs. While this was also shown to be true in $\sigma_z$ for mono-abundance populations (MAPs) in SDSS/SEGUE by \citet{2012ApJ...755..115B}, here, we extend that finding to show that mono-age, mono-\feh{} populations \citep[for which MAPs are near analogues, aside from a few caveats, see][]{2017ApJ...834...27M} show very little variation in $\sigma_z$ \emph{and} $\sigma_R$ as a function of $|z|$. It is worth noting at this point that \citet{2012ApJ...755..115B} used the isothermality of MAPs to demonstrate the precision of the SEGUE element abundances. While we do not rigourously perform a similar test here, our finding that most mono-age bins demonstrate isothermality over a large range in $|z|$ is likewise indicative that the precision in our age measurements is likely similar to the size of bins that we use \citep[given the assumptions made in][]{2012ApJ...755..115B}. 

An interesting exception to the isothermality is apparent in the younger populations shown in Figure \ref{fig:sigfzage}. Both $\sigma_z$ and $\sigma_R$ in these populations \emph{does} appear to show some variation as a function of $|z|$, such that the velocity dispersion in both directions is slightly increased at higher $|z|$. This is unlikely to be due to uncertainties in the age measurement, as the ages are well constrained at the youngest ages. A slightly increasing $\sigma_z$ as a function of $|z|$ was also seen in the lowest \afe{} MAPs in \citet{2012ApJ...755..115B}, which are likely roughly equivalent to these young populations. We return to this in Section \ref{sec:sigratio}.

Further to the generally isothermal nature of the mono-age, mono-\feh{} populations, we also find that the dependence of $\sigma_R$ and $\sigma_z$ on $R$ is generally very weak. The scale lengths of the assumed exponential dependencies are long, with median values across all age and \feh{} bins of $h_{\sigma_R} = 15_{-4}^{+11}$ kpc and $h_{\sigma_z} = 16_{-5}^{+19}$ kpc. However, the constraints on the scale lengths are relatively poor in comparison to the stronger constraints placed on the other parameters of the model. This is likely due to the fact that in any given age-\feh{} bin, the radial extent of the data is relatively small. Regardless of the poor constraints on the scale length of the $R$ dependence, we note that although we allowed the data to be fit with either increasing or decreasing $\sigma$ as a function of $R$, we find that it is best fit by a slowly decreasing profile in all cases. 

Finally, we find that the data are fit in most of the age-\feh{} bins by models where the tilt angle of the velocity ellipsoid $\alpha$ is aligned between the Galactocentric spherical and cylindrical coordinate system (i.e. we find that $\alpha_1$ lies between 0 and 1, with $\alpha_1=0$ corresponding to cylindrical and $\alpha_1=1$ corresponding to spherical), with a large uncertainty, such that most bins are consistent with $\alpha_1=0$. We compared our fits to the work of \citet{2015MNRAS.452..956B}, who found that stars between $0.5 \lesssim |z| \lesssim 2.0$ kpc were well fit by a relationship close to a spherical alignment, finding that $\alpha = 0.9 \arctan(|z|/R_{\odot})-0.01$. On directly comparing our fits with their results (plotting $\alpha(z_{1/2})$ against $z_{1/2}$), we find that the mono-age, mono-\feh{} populations at $z_{1/2} \gtrsim 0.5$ kpc agree well with \citet{2015MNRAS.452..956B}, albeit with very large uncertainties. Younger bins, at lower $z_{1/2}$, also appear to lie on the \citet{2015MNRAS.452..956B} relation. { \citet{2014MNRAS.439.1231B} found similarly consistent results for the tilt angle in the RAVE data.}

\subsection{Age - velocity dispersion relations in the high and low \afe{} discs}

\begin{figure*}
\includegraphics[width=\textwidth]{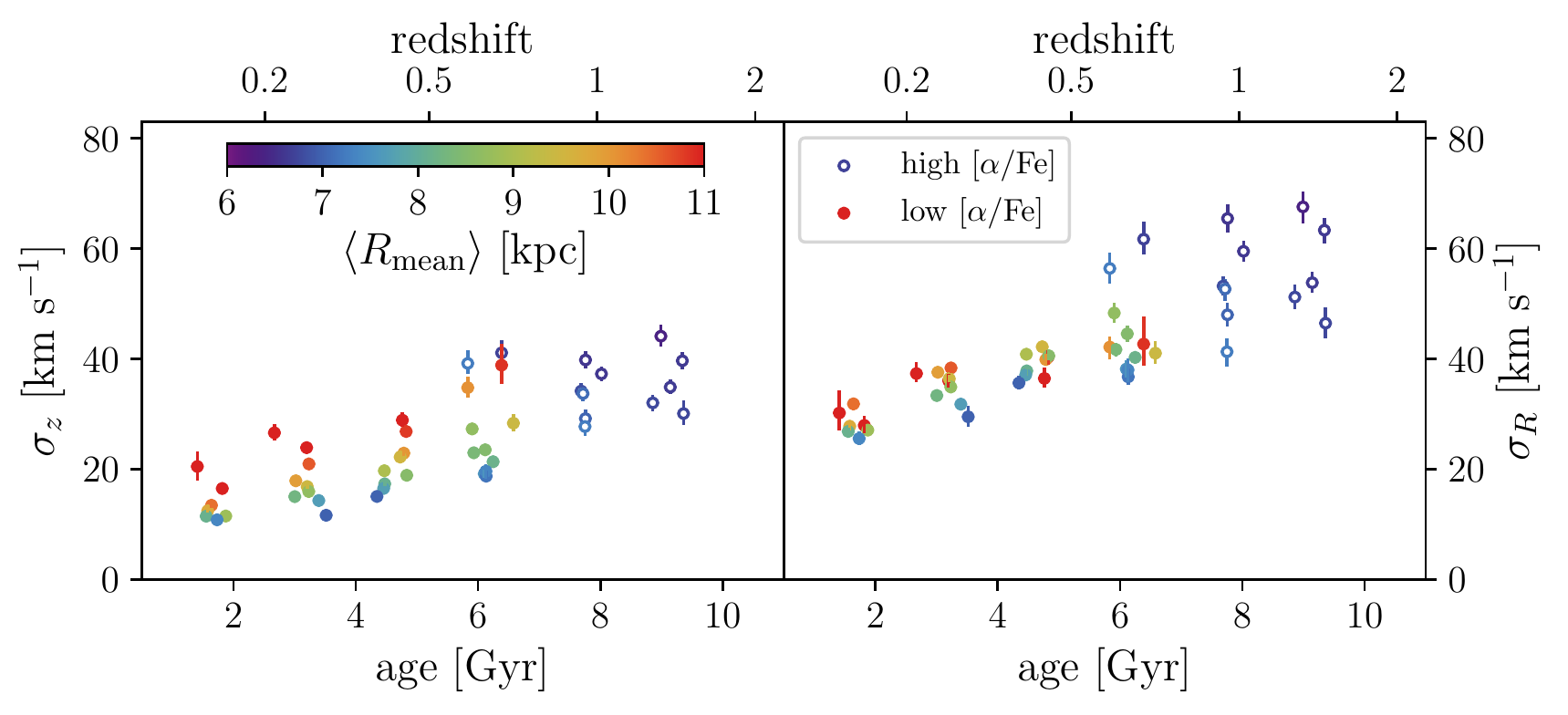}
\caption{\label{fig:sigmasage}Velocity dispersions $\sigma_R$ (\emph{right}) and $\sigma_z$ (\emph{left}) at $R_0$ and the median height $z_{1/2}(\mathrm{age,[Fe/H]})$ as a function of age in mono-age, mono-\feh{} bins. At each age, we display all the mono-\feh{} bins that have more than 200 stars and we apply a small random Gaussian jitter in age with a scale of $0.2$ Gyr to the points, to make the variation (or lack thereof) in $\sigma_{[R,z]}$ with \feh{} at fixed age more clear. The points are coloured by the median of the mean orbital radius $\langle r_{\mathrm{mean}} \rangle$ in each bin, to demonstrate the Galactocentric radius at which the stars reside. The dispersions $\sigma_R$ and $\sigma_z$ increase with age. At ages $< 6$ Gyr, there is a large spread in $\langle R_{\mathrm{mean}} \rangle$ at fixed age, where the mono-age populations with greater $\langle R_{\mathrm{mean}} \rangle$ also have higher vertical velocity dispersion $\sigma_z$, while the $\sigma_R$ is roughly the same.}
\end{figure*}

%Given that the age precision is likely to be good, or at least roughly equivalent to the bin size, and that any trends with $\sigma_R$ and $\sigma_z$ with $R$ and $|z|$ are weak, 
We now turn to examining the trends between $\sigma_R$ and $\sigma_z$ with age. Quoted and displayed values of $\sigma_{[R,z]}$ in the following are those evaluated at $R_0$ and the median height $z_{1/2}(\mathrm{age,[Fe/H]})$ which, due to the weak dependence of $\sigma_{[R,z]}$ with $z$, essentially corresponds to $\sigma_{[R,z]}(R_0)$. Figure \ref{fig:sigmasage} shows $\sigma_z$ (\emph{left}) and $\sigma_R$ (\emph{right}) against age for the mono-age, mono-\feh{} populations in the low and high \afe{} populations. At each age, we show all mono-\feh{} bins in which there were more than 200 stars. We colour the points by the median of the mean orbital radii of the stars in each bin $\langle R_\mathrm{mean} \rangle$, calculated as 
\begin{equation}
R_{\mathrm{mean}} = \frac{(r_\mathrm{peri}+r_\mathrm{ap})}{2},
\end{equation}
where $r_\mathrm{peri}$ and $r_\mathrm{ap}$ are the peri- and apocentre radius of the orbits, which are determined using the orbital parameter estimation method of \citet{2018arXiv180202592M}, as implemented in the \texttt{galpy} python package for galactic dynamics \citep{2015ApJS..216...29B}. The mono-age, mono-\feh{} populations in the high \afe{} disc are displayed as open points, whereas the low \afe{} populations are solid points.

The velocity dispersions $\sigma_z$ and $\sigma_R$ both increase roughly monotonically as a function of age, with $\sigma_z$ going from $\sim 10$ to $\sim 40\, \mathrm{km\ s^{-1}}$ and $\sigma_R$ increasing from $\sim 30$ to a maximum of $\sim 70\, \mathrm{km\ s^{-1}}$ in the oldest, high \afe{} bins. The spread in $\sigma_z$ at fixed age is greater than that in $\sigma_R$ for most age bins, but these become similar at the oldest ages. For the younger, low \afe{} populations, with $\mathrm{age} < 6$ Gyr, the bins with higher $\sigma_z$ correspond to those with the largest $\langle R_\mathrm{mean} \rangle$, indicating that these stars are typically orbiting outside the solar circle.  This indicates that $\sigma_z$ increases with $R$ at all ages. {\red Given that this is a somewhat unexpected finding, we briefly discuss its implications in Section \ref{sec:discussion}.}

The increasing $\sigma_z$ vs. $\langle R_\mathrm{mean} \rangle$  trend does not appear in the high \afe{} populations, although the spread in $\sigma_z$ is similar. For low \afe{} bins, the spread in $\sigma_R$ at fixed age is very small, such that the bins with high $\langle R_\mathrm{mean} \rangle$ have an excess of $\sigma_z$, as discussed further below. The high \afe{} populations have lower $\langle R_\mathrm{mean} \rangle$ than the low \afe{}, but have the higher $\sigma_R$ and $\sigma_z$. Importantly, the considerable spread in $\sigma_R$ at fixed age in the high \afe{} populations (and lack thereof for low \afe{} bins) is very clear.

\begin{figure*}
\includegraphics[width=\textwidth]{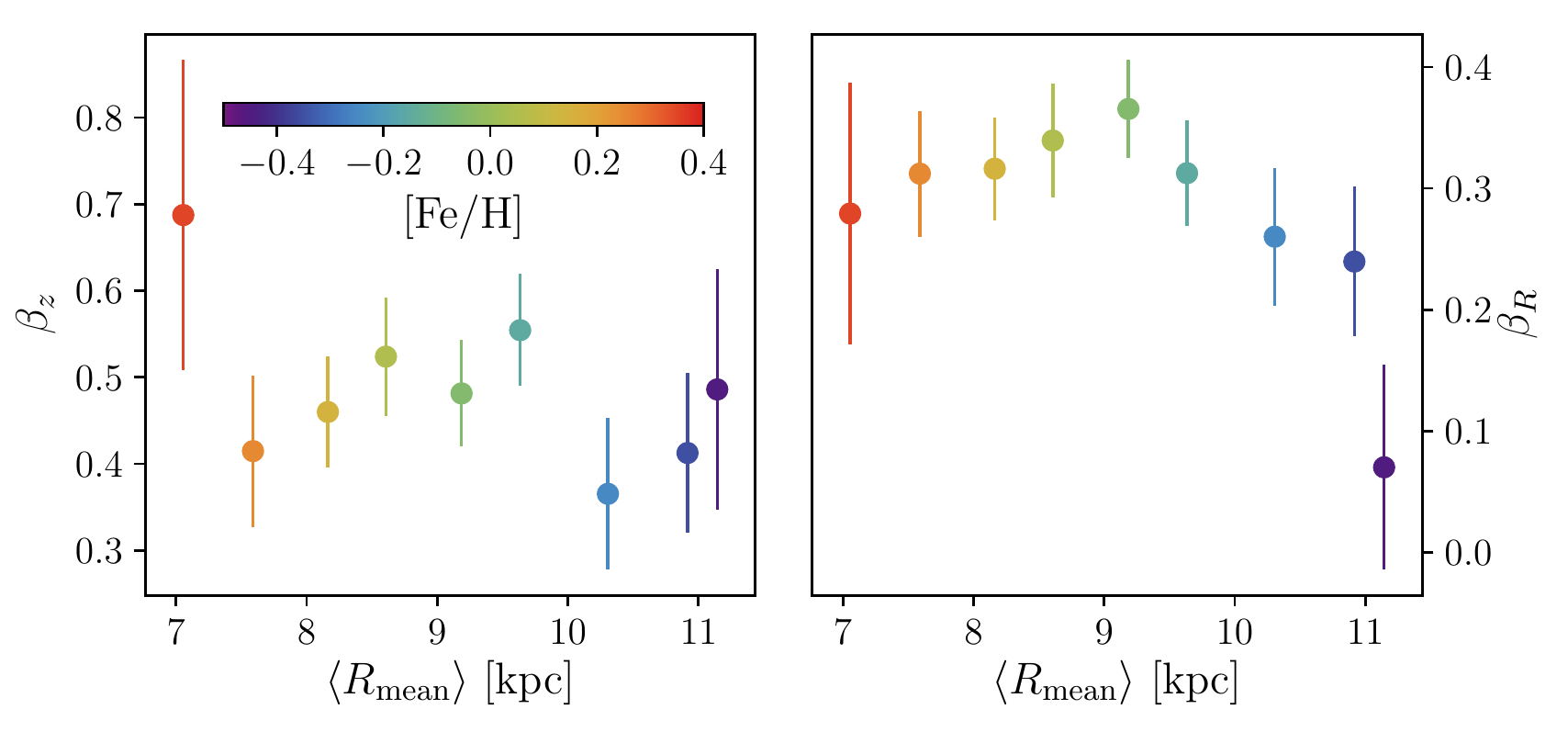}
\caption{\label{fig:betaR} The power law index of the age-velocity dispersion relations $\beta_R$ and $\beta_z$ obtained by fitting a power-law behaviour to the $\sigma_R$ and $\sigma_z$ trends as a function of age for each mono-\feh{} bin that has $>3$ mono-age bins with $N>200$ stars in the low \afe{} disc. The x-axis is  the median of the mean orbital radii in each of the mono-\feh{} bins and the points are coloured by the \feh{} of the bin. The outer disc, low \feh{} stars have flatter \emph{radial} AVRs than the stellar populations in the inner disc (with higher \feh{}). The \emph{vertical} AVRs do not change shape significantly as a function of $\langle R_{\mathrm{mean}} \rangle$}
\end{figure*}

We approximate the shape of the radial and vertical age-velocity dispersion relations (AVRs) by fitting simple power-law relationships $\sigma_{[R,z]}(\mathrm{age})\propto \mathrm{age}^{\beta_{[R,z]}}$ \citep[similarly to those fit by, e.g.][]{2016MNRAS.462.1697A,2018arXiv180803278T} to the run of $\sigma_z$ and $\sigma_R$ as a function of age in each \feh{} bin. We only fit the relationship in \feh{} bins where there are 3 or more age bins with more than 200 stars, and use a simple maximum likelihood procedure to estimate $\beta_R$ and $\beta_z$. The resulting $\beta$ for the vertical and radial AVRs as a function of $\langle R_\mathrm{mean} \rangle$ is shown for the low \afe{} mono-\feh{} bins in Figure \ref{fig:betaR}. The left panel shows that low \afe{} populations have a similarly shaped \emph{vertical} AVR as a function of \feh{} (and therefore roughly constant with $R_\mathrm{mean}$ also), such that the mean $\beta_z = 0.5 \pm 0.1$ when these populations are combined. We also fit the high \afe{} populations (but do not show these in Figure \ref{fig:betaR}). For those that were fit, we find that although less well constrained, these populations have seemingly flat vertical AVRs, with $\beta_z = 0.02^{+0.12}_{-0.03}$. We strongly emphasize, however, that age uncertainties are much larger for high \afe{} stars. Therefore, contamination between these age bins is likely and this contamination would flatten any intrinsic AVR. {  Furthermore, it is likely that the age spread in the high \afe{} stars is artificially enhanced by these uncertainties, and so the AVR for these populations has little significance.}

While $\beta_z$ is roughly constant as a function of \feh{} (and therefore $R_\mathrm{mean}$) in the low and high \afe{} discs, the shapes of the \emph{radial} AVRs for the low \afe{} populations change significantly. We find that the high \afe{} population radial AVRs are consistent with one another, and again with being nearly flat, having $\beta_R = 0.2 \pm 0.2$. However, for the low \afe{} disc populations we find that the radial AVRs \emph{do} appear to change shape with \feh{}. The values of $\beta_R$ against $\langle R_{\mathrm{mean}} \rangle$ (coloured by \feh{}) fit for the low \afe{} disc are shown in the right panel of Figure \ref{fig:betaR}. It is clear from this figure that the outer disc populations (those with the lowest \feh{}) have significantly flatter radial AVRs than those populations that reside closer to the solar radius. The index $\beta_R$ appears to increase to a maximum $\sim 0.35$ at $\langle R_\mathrm{mean} \rangle \sim 8$ kpc, and then decline slightly again at $\langle R_\mathrm{mean} \rangle$ inside the solar radius. However, within the uncertainties, all the mono-\feh{} populations with $\langle R_\mathrm{mean} \rangle$ $\lesssim 9$ kpc are roughly consistent with the same $\beta_R$. We discuss the implication of this finding in Section \ref{sec:discussion}.

\subsection{The shape of the velocity ellipsoid as a function of age}

\begin{figure}
\includegraphics[width=\columnwidth]{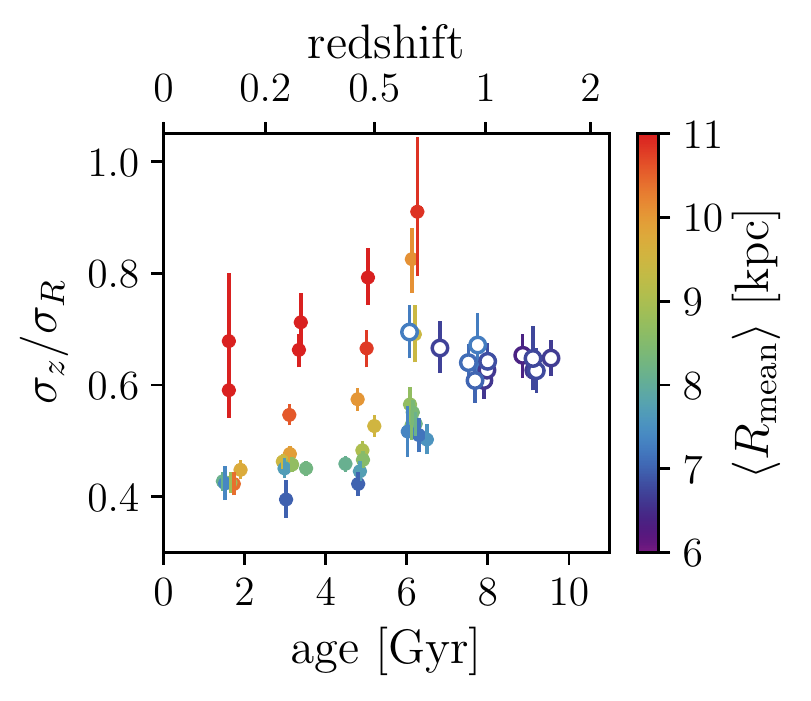}
\caption{\label{fig:sigratioage}The ratio of vertical to radial velocity dispersion $\sigma_z/\sigma_R$ in mono-age, mono-\feh{} populations as a function of age. The same jitter as in Figure \ref{fig:sigmasage} is applied to each point and the points are again coloured by the median of the mean orbital radii of the stars in each bin, $\langle r_{\mathrm{mean}} \rangle$. Bins older than $\sim 6$ Gyr have $\sigma_z/\sigma_R \sim 0.6$. At ages $< 6$ Gyr, there appear to be two separated tracks in age-$\sigma_z/\sigma_R$, where bins with stars mainly in the outer disc have $\sigma_z/\sigma_R > 0.6$, and bins with stars nearer to the solar radius have lower values. At younger ages, there seems to be a positive correlation between $\sigma_z/\sigma_R$ and age, which is roughly the same for both tracks.}
\end{figure}

To examine the kinematic structure of the disc as a function of age, in Figure \ref{fig:sigratioage} we display the axis ratio $\sigma_z/\sigma_R$ as a function of age for the mono-age, mono-\feh{} populations. As in Figure \ref{fig:sigmasage}, the points are coloured by $\langle R_\mathrm{mean} \rangle$, and high \afe{} populations are plotted as open points. In the two oldest age bins ($\gtrsim 6$ Gyr), where the high \afe{} stars reside, the relation between $\sigma_z/\sigma_R$ and age is roughly flat, and has very little scatter. Combining all the MCMC samples for these bins, we find $\sigma_z/\sigma_R = 0.64\pm{-0.04}$. {  This value is in rough agreement with the measurements of $\sigma_R/\sigma_z$ in LAMOST-TGAS stars by \citet{2018MNRAS.475.1093Y}.}

For low \afe{} populations, the relation between age and $\sigma_z/\sigma_R$ is more complex. The apparent excess of $\sigma_z$ seen in Figure \ref{fig:sigmasage} is very clear: bins at the same age with different $\langle R_\mathrm{mean} \rangle$ have different $\sigma_z/\sigma_R$, such that the mono-age, mono-\feh{} populations residing outside the solar radius have $\sigma_z/\sigma_R \gtrsim 0.7$, with many bins having higher  $\sigma_z/\sigma_R$ than the oldest, high \afe{} bins. At fixed age in the low \afe{} populations, greater values of  $\sigma_z/\sigma_R$ correspond to larger $\langle R_\mathrm{mean} \rangle$. This trend is not apparent in the high \afe{} disc, where all populations have a similar $\langle R_\mathrm{mean} \rangle$, and correspondingly similar $\sigma_z/\sigma_R$ (as discussed above). That outer disc populations are those with the highest $\sigma_z/\sigma_R$ due to an excess of $\sigma_z$ may seem unsurprising at first, given that it is well known that at young ages (and  low \afe{}), the disc is strongly flared \citep{2016ApJ...823...30B,2017arXiv170600018M} and that the disc has a significant warp in its outer regions \citep[e.g][]{2018MNRAS.481L..21P}. However, we discuss the possible effects of this warping and other disc heating agents further in Section \ref{sec:discussion}. The more novel result here is that there is an age (between $\sim 6$ and 8 Gyr ago) at which mono-age, mono-\feh{} populations exhibit a significant change in their velocity structure, which coincides with the age that roughly divides the high and low \afe{} disc populations. This may have important implications for the history of formation and assembly of the Galaxy, which we will also discuss in Section \ref{sec:discussion}.

\begin{figure*}
\includegraphics[width=\textwidth]{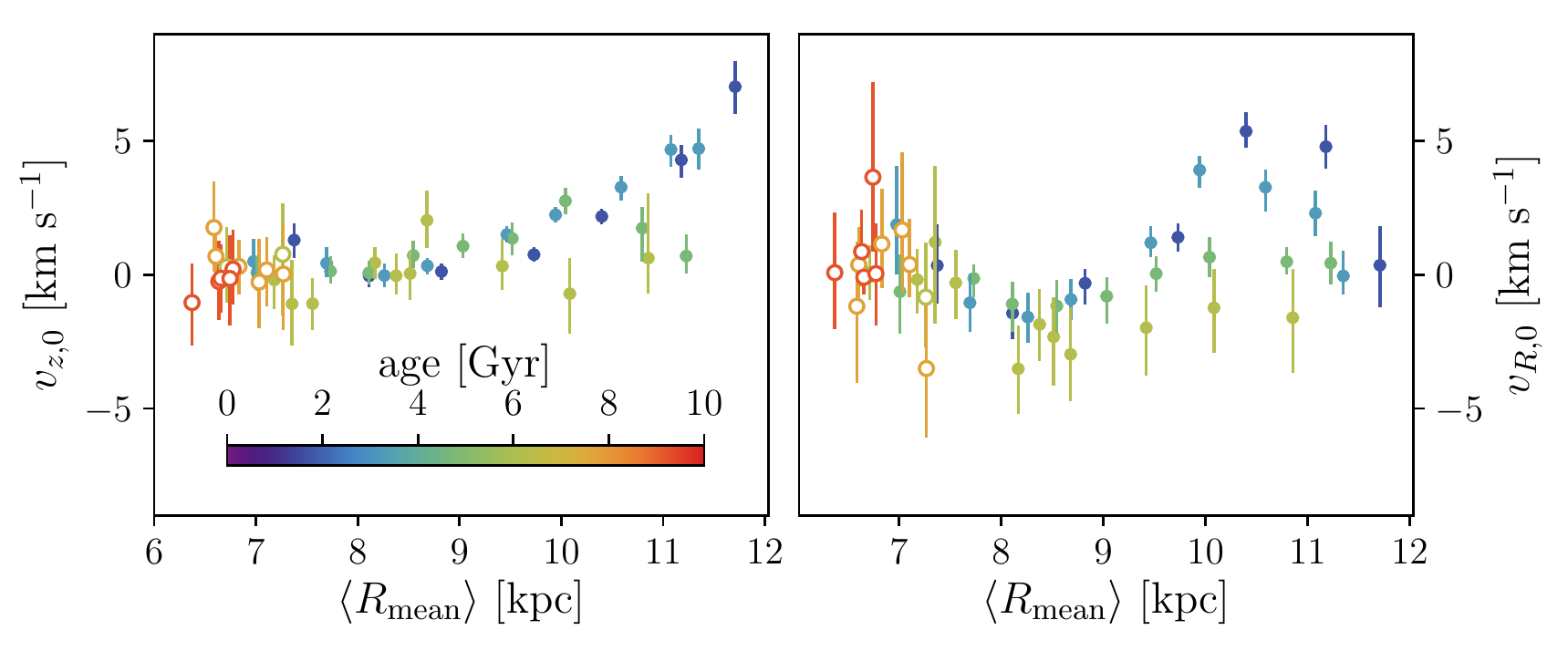}
\caption{\label{fig:vz0} The mean vertical and radial velocity $v_{z,0}$ and $v_{R,0}$ of mono-age, mono-\feh{} populations in the low (filled points) and high (open points) \afe{} disc populations as a function of the $\langle R_{\mathrm{mean}} \rangle$ of the population. The points are coloured by the age of the stars in each bin. High \afe{} populations are all consistent with $v_{z,0} = v_{R,0} = 0$ and are concentrated at low $\langle R_{\mathrm{mean}} \rangle$. Low \afe{} populations show interesting trends with $\langle R_{\mathrm{mean}} \rangle$. Populations younger than $\sim 4$ Gyr show an increasingly positive mean vertical velocity as a function of $\langle R_{\mathrm{mean}} \rangle$ and display a wave-like pattern in $v_{R,0}$ as a function of $\langle R_{\mathrm{mean}} \rangle$. Old stars in the low \afe{} disc appear to not show any significant trends in $v_{z,0}$ or $v_{R,0}$ with $\langle R_{\mathrm{mean}} \rangle$, despite their similar radial extent to the younger populations.}
\end{figure*}

\subsection{The mean velocity -- $\langle R_{\mathrm{mean}} \rangle$ relationship of populations as a function of age}
{  By allowing the mean velocities of the model velocity distributions, $v_{R,0}$ and $v_{z,0}$, to be free parameters in the fits, we can assess the degree by which the velocity distribution is subject to non-axisymmetries in the disc. We find that the general results pertaining to $\sigma_z$ and $\sigma_R$ are robust to holding the mean velocities fixed at $0\ \mathrm{km\ s^{-1}}$, and that the majority of the mono-age, mono-\feh{} populations are well fit by models with $v_{z,0} = v_{R,0} = 0$. However, we find that some mono-age, mono-\feh{} populations are better fit by models that have a significant departures from $v_{[R,z],0} = 0$. 

In Figure \ref{fig:vz0}, we show that there are clear trends between the $v_{z,0}$ and $v_{R,0}$ of mono-age, mono-\feh{} populations and the typical orbital radius $\langle R_{\mathrm{mean}} \rangle$ of the populations: young ($\lesssim 4$ Gyr), low \afe{} populations have a small but significant increasing trend of their mean vertical velocity with $\langle R_{\mathrm{mean}} \rangle$, increasing to $\sim 5\ \mathrm{km\ s^{-1}}$ for the outermost populations. These populations retain the same high $\sigma_z/\sigma_R$ values (within the uncertainties) as when fit with the fixed mean velocity model. This trend of $v_{z,0}$ with $\langle R_\mathrm{mean} \rangle$ in young low \afe{} stars is roughly consistent with the position and magnitude of the warp seen in TGAS by \citet{2018MNRAS.478.3809S} and in \emph{Gaia} DR2 by \citet{2018MNRAS.481L..21P} (their Figure 3). The scale of the warping found in our best fit models is also roughly consistent with that found in the 2MASS data \citep{2006A&A...451..515M}.

We find a different trend of the mean \emph{radial} velocity with $\langle R_{\mathrm{mean}} \rangle$ for the young, low \afe{} populations, such that $v_{R,0}$ has a wave-like pattern with increasing $\langle R_{\mathrm{mean}} \rangle$, dipping down to $\sim -3\ \mathrm{km\ s^{-1}}$ at $\langle R_{\mathrm{mean}} \rangle \sim 8.5$, before rising again to $\sim 5\ \mathrm{km\ s^{-1}}$ at the greatest $\langle R_{\mathrm{mean}} \rangle$. Interestingly, the older populations ($\sim 6$ Gyr) in the low \afe{} disc do not show any significant trends in their mean velocities with $\langle R_{\mathrm{mean}} \rangle$. Similarly, we find that all the high \afe{} populations are consistent with $v_{z,0} = v_{R,0} = 0$, although these populations are far more centrally concentrated than the older, low \afe{} populations. We further discuss this feature of the velocity distribution of the disc and it's possible bearing on its history in Section \ref{sec:discussion}}

\section{Discussion}
\label{sec:discussion}

We have shown that when regions beyond the solar vicinity are considered, the relationship between the age and the different components of the velocity dispersion of stars in the disc is complex, and co-dependent on the stars' mean orbital positions and their element abundances. In the following section, we contextualise our results within the existing work on the velocity structure of the disc and discuss the constraints that these results place on models for the formation and evolution of the Galactic disc.

\subsection{The increasing trend of $\sigma_{z}$ with $R_\mathrm{mean}$ in the low \afe{} disc}

{\red Given that our finding that $\sigma_z$ shows a clear increase with $R_\mathrm{mean}$ at all ages is somewhat surprising, as the expectation is that discs are most likely to show a \emph{decreasing} trend, we briefly discuss here some possible explanations for this finding, and present some comparison to other observational results.

For a disc in equilibrium:
\begin{equation}
    \Sigma(R) \propto \frac{\sigma_z^{2}}{h_z}
\end{equation}
where $\Sigma(R)$ is the disc surface density, and $h_z$ is the disc scale height. It follows that if $h_z$ is not a function of $R$, then $\sigma_z$ should trace $\Sigma(R)$ and therefore be a declining function of $R$. This relationship between $\sigma_z$ and $\Sigma(R)$ is commonly used to measure the surface density of external galaxy discs, making an assumption of a radially constant $h_z$, which is difficult to measure in face-on galaxies where $\sigma_z$ is more readily measurable. It is commonly seen in these studies that $\sigma_z$ declines with $R$ \citep[see, e.g.][]{2013A&A...557A.130M,2018MNRAS.476.1909A}. The assumption of constant $h_z(R)$ is usually based on observations of edge-on spiral galaxies with small bulges \citep[e.g.][]{2006AJ....131..226Y}, likely smaller than that of the Milky Way, whose bulge and bar extends out to ~3-4 kpc \citep[e.g.][]{2015MNRAS.450.4050W}. Lastly, when comparing these results with those from external galaxies, where stellar populations are not divided by age or abundances, it is important to note that we \emph{do} make such divisions here. \citet{2015ApJ...804L...9M} showed that measured trends in $h_z$ can be quite different between mono-abundance populations and the total population, and it is possible that these effects also play a role in the $\sigma_z$-$R$ relationship of mono-age, mono-\feh{} populations.

Aside from these points, the Milky Way does appear from these results, however, to diverge from external galaxies in this respect. Flattening and a slight increase in $\sigma_z$ with $R$ outside the solar radius was also found by \citet{2018MNRAS.481.4093S}. It is known that mono-age populations in the disc flare considerably \citep[e.g.][]{2016ApJ...823...30B,2017arXiv170600018M}. Since $h_z$ then depends on $R$, $\sigma_z$ can stay constant or increase with $R$. The existence of the extended bar in the Milky Way may also affect this relationship, and the specific details of its structure and dynamics, and interaction with disc stars, is as yet not well determined. Our finding is perhaps more logical still, given that the disc is likely not in equilibrium, but being perturbed by external forces such as interactions with satellites.}

\subsection{Heating agents and the velocity dispersion history throughout the disc}
\label{sec:heatingagents}
Our understanding of heating in the Milky Way disc has thus far largely been informed by observational constraints that were confined relatively near to the Sun. Our data set extends well beyond the solar vicinity, out to $R \gtrsim 12$ kpc in the outer disc and into the inner disc down to $R \lesssim 5$ kpc. This allows us to build a picture of heating and its connection to Galactic evolution over the extent of the disc. While a more robust understanding of these results will likely only be afforded by detailed numerical simulations, we discuss the implications of our findings using the extensive body of existing work on this topic. {  We attempt to understand trends in our results from the context of the more classical results on this topic, which, although outdated in some regards by numerical simulations, provide some key insights into the processes at hand. Providing reference to the most up-to-date theoretical work on the topic \citep[using numerical simulations, e.g. those examined in][]{2016MNRAS.462.1697A,2016MNRAS.459.3326A,2017MNRAS.470.3685A,2017MNRAS.470.2113A} allows us to consolidate these findings into a coherent picture of the heating processes which have shaped the Milky Way disc.}

The majority of studies that aim to understand the history of the disc from its velocity structure concentrate on two main observables, which are those that we have measured and discussed in Section \ref{sec:results}: a) the shape of the age-velocity dispersion relations, usually characterised by a power law with index $\beta$, and b) the ratio of the vertical to radial velocity dispersion $\sigma_z/\sigma_R$. Along with these observables, two main heating agents are generally considered: scattering by small, localized perturbations in the potential such as GMCs \citep[e.g.][]{1951ApJ...114..385S,1953ApJ...118..106S,1984MNRAS.208..687L} and scattering by spiral arms \citep[e.g.][]{1984ApJ...282...61S,1985ApJ...292...79C}. The general picture that has emerged is that GMCs are responsible for much of the heating in the vertical direction, whereas spiral arms act to heat the disc radially. Spirals contribute little to the vertical heating because the scale lengths of the spiral irregularities ought to be much longer than the vertical scale of the stellar disc \citep[e.g.][]{1990MNRAS.245..305J}.  

\subsubsection{Velocity ellipsoid shape}

Analytical formulations of the (vertical) heating due to GMCs by \citet{1984MNRAS.208..687L} showed that the velocity dispersion should increase as $\sigma_z \propto \tau^{0.25}$, and that the axis ratio $\sigma_z/\sigma_R$ should approach a value of $\sim 0.8$. Later work, using more sophisticated calculations, found that the axis ratio resulting from GMC heating is more likely somewhere between $\sim 0.5$ to $0.6$, but found a similar power-law index of $\sim 0.3$ \citep{1993MNRAS.263..875I,2002MNRAS.337..731H}. We find that all high \afe{} populations are consistent with having $\sigma_z/\sigma_R = 0.64\pm 0.04$, in rough agreement with this revised value for GMC heating { (we return to the AVR later in Section \ref{sec:avrs})}. %However, the near-flat AVRs of the high \afe{} populations are in contention with predictions for heating by GMCs. 
In addition to this, it is interesting to note that all high \afe{} populations have similar, low $\langle R_{\mathrm{mean}} \rangle$, but large variations in velocity dispersion with \feh{} in a fixed age bin. That these populations were presumably formed at similar times and have different radial and vertical velocity dispersions but a similar ratio of these may indicate that these stars in fact formed in a turbulent ISM {\red \citep[external galaxy observations between redshift 0.7 and 2.3 indicate that the ISM is indeed turbulent at these early times, e.g.][]{2015ApJ...799..209W}}, and have simply retained their kinematics. Understanding this will, however, require further modelling of the formation of these stellar populations in self-consistent cosmological simulations.

In the low \afe{} disc, the ratio $\sigma_z/\sigma_R$ varies strongly as a function of \feh{} (and therefore $\langle R_\mathrm{mean} \rangle$) and age, with many of the low \feh{} (outer disc) populations having $\sigma_z/\sigma_R$ exceeding that predicted by GMC models, and the higher \feh{} (inner disc) populations falling well below the predicted value. The fact that each \feh{} bin in the low \afe{} disc also shows an increasing trend between $\sigma_z/\sigma_R$ and age also indicates that the kinematics of this population cannot be explained by GMC heating alone, because models predict a flat $\sigma_z/\sigma_R$ with age if a disc is heated by a single stationary process \citep[this is seen, for example, in the work of][]{1990MNRAS.245..305J}. {  Predictions of $\sigma_z/\sigma_R$-age relations from more recent numerical simulations which include a changing mass fraction of GMCs over time show clearly similarly increasing trends to those seen in our results \citep[][]{2016MNRAS.462.1697A}. In those simulations, GMC scattering is important for the early disc, and slowly gives way to spiral arm heating, as the disc grows in mass. Our result that $\sigma_z/\sigma_R$ increases with age in the low \afe{} disc is clearly in-line with this picture of heating by non-stationary processes. A similar scenario was proposed also by \citet{2018MNRAS.475.1093Y}, whose results on the velocity ellipsoid shape are consistent with those presented here.}

\citet{1990MNRAS.245..305J} showed that increasing the importance of spirals over GMCs \emph{reduced} $\sigma_z/\sigma_R$, such that models that have the most dominant spiral arm perturbations have the lowest $\sigma_z/\sigma_R$. {  This is also borne out in numerical simulations \citep[e.g.][]{2016MNRAS.462.1697A}, where spirals heat generally only in the plane (increasing $\sigma_R$), whereas GMCs heat both in-plane and vertically. By this logic $\sigma_z/\sigma_R$ becomes a good measure of the relative importance of these processes in shaping the disc, but it is important to bear in mind also the external effect of satellite interactions which should also perturb and heat the disc \citep[such interactions are not included, for example, in][]{2016MNRAS.462.1697A}}.  We find that in the low \afe{} disc, the outermost disc populations (at low \feh{}) have the highest $\sigma_z/\sigma_R$ at fixed age. This is consistent with the picture of declining importance of spiral arm heating with Galactocentric radius. However, an important finding here is that the highest values of $\sigma_z/\sigma_R$ in the outer disc far exceed that expected from GMC or spiral heating. {  The resulting $\sigma_z/\sigma_R$ for the low \afe{} populations in the outer disc also exceeds that found in many of the idealised simulations of \citet{2016MNRAS.462.1697A}}.  It is logical to assume that the number of GMCs in the outer disc is less than that in the inner disc \cite[indeed, this is readily observed in the Milky Way, e.g.][]{2017ApJ...834...57M}, and likely even less at earlier times, while the disc was still growing. Therefore, it seems unlikely that GMCs or spiral arms could have heated these outer disc populations to their very high velocity dispersion. {  It is notable that this increasing vertical velocity dispersion with R is also realised in the results of \citet{2018MNRAS.481.4093S} (their Figure 9).} Fully understanding this excess of vertical velocity dispersion will almost certainly require additional modelling of the spatial density structure of the disc, which can then be compared with the kinematic modelling performed here.

\subsubsection{The age-velocity dispersion relation}
\label{sec:avrs}
It is also notable that we find that $\sigma_z \propto \tau^{0.5}$ for the low \afe{} disc populations, in general agreement with previous observational work \citep[e.g.][]{1977A&A....60..263W,2007MNRAS.380.1348S,2008A&A...480...91S} and in disagreement the classical theoretical value for GMC heating. {  This discrepancy has been widely noted in the literature, and was cast by \citet{2016MNRAS.462.1697A} as arising from the discrepancy between AVRs and true \emph{heating histories}, e.g. the evolution of velocity dispersion in co-eval populations which generates the AVR. They find that heating histories (described by a power law index $\tilde{\beta}_R$), which are mainly set in the few Gyr after stellar birth, tend to depend on the birth time of stars itself, such that stars born at different times in the Galaxy's history attain different kinematics due to the changing nature of perturbers with time. This leads to a discrepancy in the AVR from simple analytical values, which assume heating to be a stationary process.} 

% \citet{1990MNRAS.245..305J} showed that the inclusion of perturbations by spiral arms of increasing relative importance to that of heating by GMCs acts to increase the power-law index of the radial AVR to $\beta_R\sim0.4$ relative to their model with GMCs only, which had $\beta_R \sim \beta_z \sim 0.25$ (These models could also only slightly increase the vertical AVR slope to $\beta_z \sim 0.3$). \citet{2004MNRAS.350..627D} found a wider range of possible $\beta_R$ between $\sim 0.25$ and $\sim 0.5$ using direct numerical orbit integrations for a variety of different spiral-arms models.
%These spiral-arm predictions provide an interesting insight into the possible origin of some of the trends that we observe in the APOGEE-\emph{Gaia} data. 
{  Considering radial heating, the simulations of \citet{2016MNRAS.462.1697A} predict that $\beta_R$ should vary with Galactocentric radius, and also predict that the heating histories of younger populations should have greater $\tilde{\beta_R}$, likely due to the fact that in their simulations the GMC heating gives over to spiral arm perturbations \citep[this was also predicted by][although they failed to predict correct indices, for a variety of reasons]{1990MNRAS.245..305J}.}  We find that $\beta_R$ shows a clear trend with $\langle R_\mathrm{mean} \rangle$ in the low \afe{} disc, such that for the outer disc stars $\beta_R \sim 0.15$. As $\langle R_\mathrm{mean} \rangle$ decreases to $\lesssim 10$ kpc, $\beta_R$ rapidly increases to values between $\sim 0.3$ and $0.4$. The decreased $\beta_R$ in the outer disc stellar populations may be a manifestation of the waning influence of spiral arm heating on the kinematics of low \afe{} stellar populations that reside in the outermost part of the disc. If spiral arm heating is acting strongly inside $\sim 10$ kpc, peaking at $8 < \langle R_{\mathrm{mean}} \rangle < 9$ kpc, then this may also suggest that radial migration of stars is most efficient in this region of the disc. We also find, at relatively low significance, that $\beta_R$ increases slightly between the innermost populations and those at $\sim 8$ kpc.  %\ted{Very important point, making this note to make sure we emphasize this in conclusion}

\subsection{Warping of the disc as a heating agent}

In the absence of extensive GMC or spiral arm heating, there are few remaining possible causes for the large velocity dispersions that we observe in the outer disc. One interesting possibility is heating from warping of the Galactic disc, caused by satellite interactions \citep[e.g.][]{2014ApJ...789...90K}, bending instabilities \citep[e.g][]{2017A&A...597A.103K}, or misalignment of the angular momenta of the disc and dark halo \citep[e.g.][]{1999MNRAS.303L...7J,1999ApJ...513L.107D}, to name a few proposed processes. Such a warping of the Milky Way disc is well documented \citep[e.g. ][]{2006A&A...451..515M,2018MNRAS.478.3809S} and has been recently confirmed in the \emph{Gaia} DR2 data \citep{2018MNRAS.481L..21P}. Modelling of such bending modes has suggested that this process may act to increase $\sigma_z$ \citep[as shown in the early work of][]{1969ApJ...155..747H}, therefore increasing $\sigma_z/\sigma_R$ \citep[e.g.][]{2014MNRAS.443.2452M}. \citet{2017A&A...597A.103K} suggested heating by these processes can cause $\sigma_z/\sigma_R$ to approach values as high as unity. Therefore, it is at least possible that these stars have been heated by warping or bending of the disc.

That the amplitude of warping in the low \afe{} disc is stronger at low ages may suggest that the warp is likely stronger in the star-forming gas disc, causing more recently formed stars to retain the warp kinematics more readily. This is somewhat consistent with the finding of \citet{2017arXiv170600018M} that young stars in the low \afe{} disc are those which have a stronger flare in the outer parts of the disc. The warping present in the kinematics combined with the flattened radial AVR and excess $\sigma_z/\sigma_R$ of the outer disc populations, strongly suggests that the kinematics of these populations is affected by the warping of the Galactic disc. 

\subsection{Non-isothermality in young populations}
\label{sec:sigratio}
\begin{figure*}
\includegraphics[width=\textwidth]{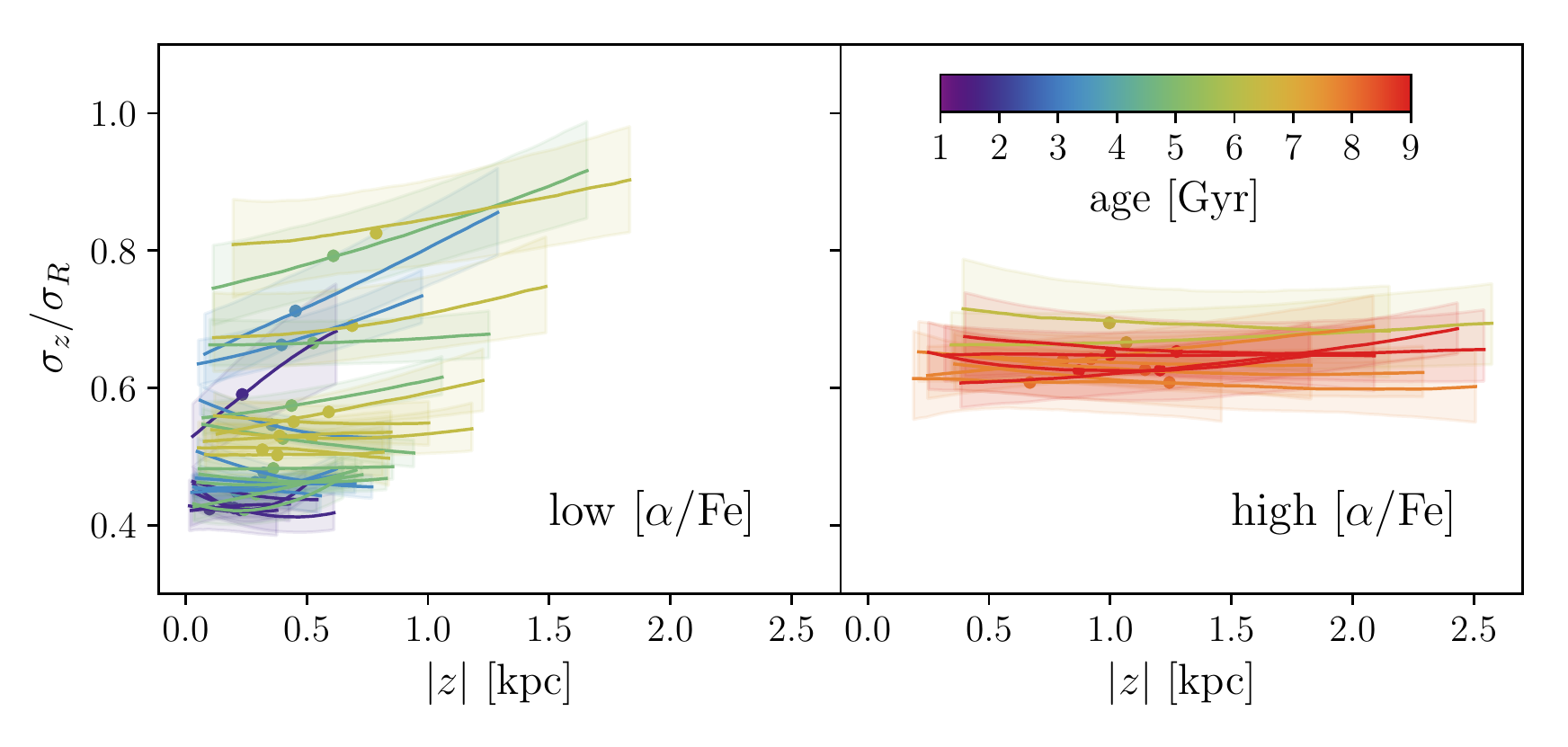}
\caption{\label{fig:sigratioprof} The profile of $\sigma_z/\sigma_R$ with $|z|$ at $R_{0}$ for mono-age, mono-\feh{} populations in the low and high \afe{} disc. Each line represents the best fit profile for that mono-age, mono-\feh{} bin, with the coloured band indicating the 95\% confidence interval. The colour indicates the age of the mono-age, mono-\feh{} population. The $\sigma_z/\sigma_R$ of young, low \afe{} populations show a slight, but significant trend with $|z|$, increasing at greater $|z|$. The decreased $\sigma_z/\sigma_R$ at lower $|z|$ is indicative that spiral heating is acting more strongly in the plane. The high \afe{} populations have very flat $\sigma_z/\sigma_R$ with $|z|$, further indicating that these populations are kinematically distinct from the low \afe{} population and that their velocity dispersion is unlikely to be due to perturbers confined to the midplane.}
\end{figure*}

We saw in Section \ref{sec:dispprofile} that the vertical dependence of the radial and vertical kinematics departs from isothermality for the youngest populations in the disc. This manifests itself in the $\sigma_{[R,z]}$ profiles as a function of $|z|$ as an increasing trend with increased height above the midplane. This feature was first hinted at at low significance in the dependence of $\sigma_z$ on $z$ for low \afe{} populations by \citet{2012ApJ...755..115B}, but is now well established from our analysis in this paper. We now investigate this further by also studying the shape of the velocity ellipsoid $\sigma_z/\sigma_R$ (as shown in Figure \ref{fig:sigratioage}) as a function of $|z|$. These profiles for mono-age, mono-\feh{} populations in the low and high \afe{} disc are shown in Figure \ref{fig:sigratioprof}.

It is immediately clear from Figure \ref{fig:sigratioprof} that the younger, low \afe{} $\sigma_z/\sigma_R$ profiles increase with $|z|$, whereas the high \afe{}, old population profiles are flat over a large range of $|z|$. Following from the discussion of the work of, e.g., \citet{2016MNRAS.462.1697A} in Section \ref{sec:heatingagents}, a decreased $\sigma_z/\sigma_R$ is likely a signature of stronger heating by spiral structure. This presents a compelling explanation for the departure from isothermality of these young populations, and may also indicate the differing timescale of these heating agents. For example, if spirals can radially heat the stellar population quickly, this would reduce the $\sigma_z/\sigma_R$ in the plane for young populations. As GMCs likely redirect this radial heating into vertical heat more slowly than the spirals heat the disc, then one might expect also a trend in $\sigma_z$ with $|z|$ for the younger populations, as the stars which have been vertically heated are at this point more likely to be above the plane. For older ages, this trend becomes erased as the populations approach isothermality due to prolonged exposure to this combination of heating agents. It is, however, apparent in Figure \ref{fig:sigratioprof} that some of the older low \afe{} populations still show an increasing $\sigma_z/\sigma_R$ profile. As these populations are those residing in the outermost regions of the disc, we propose that this may be due again to the waning effect of the spiral structure, reducing the flattening at old ages of these profiles. %\ted{I don't think the figure quite bares this out as currently written: the oldest, low-alpha populations still have a significant increase in sigmaz/sigmaR. Perhaps this is because these are the outer disc ones and the flattening at older ages is for the inner disc ones?}

That such trends are not seen at all in the high \afe{} populations further suggests a distinct evolutionary history for these populations, which we will discuss further in the following section.

\subsection{Kinematic distinction of the high and low \afe{} discs}

Our results demonstrate that there are many kinematic differences between the high and low \afe{} discs. Not only do these populations differ in their velocity dispersions $\sigma_R$ and $\sigma_z$, but they appear to have an entirely different kinematic structure, displaying different trends in their kinematics with age, \feh{}, and position in the Galaxy. 

 It is well known from previous work that the high \afe{} disc populations are old, have the lowest $\langle R_\mathrm{mean} \rangle$, and the highest velocity dispersions. In this paper, we have determined that the high \afe{} disc populations are furthermore characterised by near-flat radial and vertical AVRs and that they are consistent with a single $\sigma_z/\sigma_R = 0.64\pm 0.04$, which does not vary with age, \feh{}, or $|z|$. These trends are markedly different from those displayed by the younger, low \afe{} populations that show complex trends in $\sigma_z/\sigma_R$ with age and \feh{}, and have increasing values of this ratio with $|z|$. High \afe{} stellar populations appear to be more centrally concentrated, with smaller scale lengths and larger scale height than their low \afe{} counterparts \citep{2012ApJ...746..149C,2012ApJ...753..148B,2016ApJ...823...30B,2017arXiv170600018M}, and have the smallest $\langle R_\mathrm{mean} \rangle$ of all the mono-age mono-\feh{} populations. A centrally concentrated and kinematically distinct high \afe{} component is a key prediction in models which invoke an early, rapid formation of the high \afe{} population, induced by rapid gas inflow or high frequency gas-rich mergers \citep{1997ApJ...477..765C,2001ApJ...554.1044C,2004ApJ...612..894B,2018MNRAS.477.5072M}. The slight discontinuity in age vs. velocity dispersion between the high and low \afe{} discs (seen in Figure \ref{fig:sigmasage}) is also suggestive of a different, more abrupt origin of the high \afe{} populations than that predicted by, e.g. \citet{2009MNRAS.396..203S,2009MNRAS.399.1145S}. {  It was also predicted by \citet{2016MNRAS.459.3326A} that thick disc components do not arise during quiescent disc growth.}

The low \afe{} populations are extended in $R$, and have a rich kinematic structure in age and \feh{}, consistent with a gradual formation over a longer timescale, and a subsequent evolution that was likely affected by spiral structure, GMCs, and satellite interactions in ways which strongly depend on position in the Galaxy. As discussed in Section \ref{sec:heatingagents}, we find that the low \afe{} populations have vertical AVRs that agree well with those measured previously, where $\sigma_z \propto \tau^{0.5}$, and disagree with theoretical predictions that $\sigma_z \propto \tau^{0.25}$ for GMC and spiral heating. The \citet{2009MNRAS.396..203S,2009MNRAS.399.1145S} model suggests that the increased slope of the AVR may be due to biases from high $\sigma_z$, old, high \afe{} stars which have migrated to the solar radius. This bias should therefore be removed when fitting the AVR of \emph{only} low \afe{} stars. We do not find that the AVR slope is decreased when fitting these populations separately. As pointed out in \citet{2018arXiv180803278T}, the AVR may be steepened by invoking a larger number of scatterers (GMCs) in the past. This, contrasted with the relatively flat vertical AVR of the high \afe{} populations, further suggests that these populations had very different origins and subsequent evolutionary histories. The emergent picture then from the kinematics of the low and high \afe{} disc components is that they are quite different, and therefore likely formed and evolved differently and certainly over very different timescales. 

%The general picture emerging from this discussion is that the high and low \afe{} discs appear to be distinct not only in chemistry, but in their kinematics and spatial structure. This may seem contradictory to the contention that the vertical structure of the disc is smooth as a function of \afe{} \citep{2012ApJ...751..131B,2016ApJ...823...30B} and age \citep{2017arXiv170600018M}. However, we emphasise that these findings are in fact complementary, but demand a re-assessment of the vertical continuity between the high and low \afe{} disc in the context of the emerging new picture of the Galaxy \citep[a useful schematic of which can be found in Figure 1 of][]{2018arXiv180902658B}. 

\section{Summary and Conclusions}
\label{sec:conclusion}
In this paper, we have performed a detailed dissection of the Milky Way's low and high \afe{} disc kinematics as a function of \feh{} and age. We derive a new set of ages for APOGEE DR14 red giant branch stars by applying a Bayesian Convolutional Neural Network (BCNN) model, trained on stars with APOGEE spectra and asteroseismic ages from the APOKASC catalogue. The improved age precision, alongside the exquisite abundances afforded by APOGEE allows for a new view of the Galactic disc in age-\feh{}-\afe{} space. The large crossover with the \emph{Gaia} DR2 catalogue affords us an unparalleled catalogue of stars with measured ages, abundances, positions and kinematics with which to place constraints on models for the formation and evolution of the Galaxy. Here, we have focused on determining the kinematic structure of the disc in age-\feh-\afe{} space. Our main findings can be summarised as follows:
\begin{itemize}
\item{The radial and vertical velocity dispersions of mono-age, mono-\feh{} populations in the high and low \afe{} discs generally show little variation with $|z|$---i.e., most are close to isothermal---and have slowly decreasing exponential dependencies on $R$. The youngest mono-age populations \emph{do} however display a departure from isothermality as they slightly increase in $\sigma_R$ and $\sigma_z$ with $|z|$ (see Figures \ref{fig:sigfzage} and \ref{fig:sigratioprof}).}
\item{The tilt of the velocity ellipsoid is consistent across all mono-age, mono-\feh{} populations with alignment with the Galactocentric spherical coordinate system, but with very large uncertainties, i.e. the velocity ellipsoid points toward the Galactic centre. We find our fits of the tilt angle as a function of $z$ to be in good agreement with the findings of, e.g. \citet{2015MNRAS.452..956B} and \citet{2014MNRAS.439.1231B}, confirming that radial and vertical motions of stars in mono-age, mono-\feh{} groups are coupled.}
\item{In the age bins considered, the radial and vertical velocity dispersion increases roughly monotonically with age (Figure \ref{fig:sigmasage}). High \afe{} mono-age, mono-\feh{} populations show a spread at fixed age in both $\sigma_R$ and $\sigma_z$, whereas the low \afe{} populations only have a significant spread in $\sigma_z$, which is correlated with the mean orbital radius of the populations.}
\item{The shape of the vertical age-velocity dispersion relation does not vary between mono-age, mono-\feh{} populations in the low and high \afe{} disc. We find that $\beta_z = 0.5\pm 0.1$ for the low \afe{} disc, in general agreement with previous studies \citep[e.g][]{1977A&A....60..263W,2007MNRAS.380.1348S,2008A&A...480...91S} and the recent \emph{Gaia} DR2 work by \citet{2018arXiv180803278T}. We find $\beta_z = 0.02^{+0.12}_{-0.03}$ (consistent with a flat AVR) for the high \afe{} populations {  albeit at low significance, due to the large uncertainties on age in this regime.}}
\item{The \emph{radial} age-velocity dispersion relation changes shape as a function of \feh{} for the low \afe{} disc populations, with the lowest \feh{} populations having flatter radial AVRs (Figure \ref{fig:betaR}). These populations have the largest $\langle R_\mathrm{mean} \rangle$, suggesting that the outer disc has undergone a different history of heating to the inner disc. We tentatively attribute this to the declining effect of spiral arm heating with Galactocentric radius, and the increasing effect of external heating agents (e.g. satellite mergers) on the Galactic disc.}
\item{The shape of the velocity ellipsoid, defined by the axis ratio $\sigma_z/\sigma_R$, varies in a complex manner with age, \feh{} and \afe{} (See Figure \ref{fig:sigratioage}). High \afe{} populations have a constant $\sigma_z/\sigma_R = 0.64\pm 0.04$, which does not vary with age or \feh{}. The low \afe{} populations' $\sigma_z/\sigma_R$ display clear variations with age, and \feh{}, which is manifest also as a trend between the kinematics and their mean orbital radius $\langle R_\mathrm{mean} \rangle$. At each age, outer disc (low \feh{}) populations have the highest $\sigma_z/\sigma_R$. This increased $\sigma_z/\sigma_R$ is due to the `excess' $\sigma_z$ in the outer disc, in combination with the flattened radial AVR of these populations, reinforcing the notion that the outer disc is likely to have been heated by some process other than GMC scattering or spiral arm heating, which predict smaller $\sigma_z/\sigma_R$.}
\item{Additionally, we fit models where the mean velocity in the vertical and radial direction is allowed to vary freely. In general, mono-age, mono-\feh{} populations are well fit by models with $v_{z,0}=v_{R,0}=0$. But we find that the young ($\lesssim 4$ Gyr) low \afe{} populations with larger $\langle R_\mathrm{mean} \rangle$ seem to be better fit by models with a slightly positive $v_{[R,z],0}$, and have a wave-like trend in $v_{R,0}$ with $\langle R_\mathrm{mean} \rangle$ (See Figure \ref{fig:vz0}). Old stars in the low \afe{} disc do not appear to be part of this warping. We contend that these shifts are likely due to the Galactic warp, which may be partially responsible for the excess heating in the outer disc.}
\end{itemize}

%\ted{Maybe add a paragraph summarising the interpretation in Sec. 5.1?}
Our main findings place novel constraints on the history of dynamical heating across the disc. The finding that the velocity ellipsoid shape is strongly dependent on Galactocentric radius for low \afe{} populations, and that the AVR becomes flatter at greater radii is strongly suggestive that the Milky Way disc has been heated by external perturbers. We also propose that the changing AVR shape with $\langle R_{\mathrm{mean}} \rangle$ provides insight into where heating from spiral structure becomes most important, and may indicate where radial migration is most prevalent in the disc (for example, our results suggest that radial migration should be most efficient at $8 < R < 9$ kpc; see Section \ref{sec:heatingagents}).  We also find that the high and low \afe{} disc components appear to be kinematically distinct, reinforcing the notion that these structures formed differently.

These results pose many new constraints which must be considered in any model which attempts to explain the kinematics of the Milky Way disc. Importantly, they strongly suggest that external effects may play an important role in setting the kinematic structure of the outer disc. Furthermore, our results point to the need for the inclusion of non-axisymmetries in any modelling, as the excess heating seen in the data may be caused by the induced warping of the disc, and the kinematics of the low \afe{} population may be affected by perturbations from spiral structure. Indeed, our results also suggest that fitting the observed disc kinematics with axisymmetric models (as we do here) may be limiting the inferences that can be made from the data.

Particularly, these results make clear the need for self-consistent modelling using large volume numerical simulations in a cosmological context, which provide statistical samples of galaxies evolving in different environments and with different assembly histories, but also where the stellar dynamics and kinematics are adequately resolved. In this regard, the data are somewhat leading the theory, as simulations of large volumes tend to poorly resolve (or not model) the cold gas phases required to form stars in disc galaxies with realistic kinematics and subsequently realistic dynamical heating processes. However, the most recent zoom-in simulations are now beginning to resolve colder gas phases, and self consistently model heating from GMCs, for example \citep[e.g.][]{2016ApJ...827L..23W,2017MNRAS.467..179G}. An advantage of using such simulations is that non-axisymmetries are also modelled self-consistently, allowing the combined effects of bars, spiral arms, warps and waves on the galaxy kinematics to be assessed.

%We have presented results derived from a large data set for Milky Way stars, whose dimensionality is far greater than six-dimensional phase-space. This data set has great potential for further insights into the formation and evolution of the Galaxy. In particular, detailed dynamical modelling which can account for the now readily observed non-axisymmetric features of the disc and which does not make the (likely flawed) assumption of dynamical equilibrium will be essential for the future of Galactic archaeology. 

% \begin{itemize}
% \item Comparison with Aumer and Binney work (also Jenkins)
% \item Compare to Ting and Rix paper - reinforce importance of division in \afe{} and the modelling of sigma R.
% \item Discuss importance of the change in velocity structure, coincidence with known events in MW history
% \item context alongside the spatial structure work 
% \item potential age systematics problems? also systematics in Gaia??
%\end{itemize}

\section*{Acknowledgements}

The authors thank the anonymous referee for a useful report. We also thank Ivan Minchev, Jerry Sellwood, Chao Liu and Michael Hayden for helpful discussions on the originally submitted manuscript. We also thank Leandro Silva and Martin Smith for highlighting necessary typographic corrections. JTM is funded by an STFC studentship, and is grateful for funding from the Royal Astronomical Society and the Dunlap Institute for Astronomy and Astrophysics at the University of Toronto and thanks them for their hospitality during an extended visit while preparing this work. JTM and AM acknowledge support from the ERC Consolidator Grant funding scheme (project ASTEROCHRONOMETRY, G.A. n. 772293). JB acknowledges the support of the Natural Sciences and Engineering Research Council of Canada (NSERC), funding reference number RGPIN-2015-05235, and from an Alfred P. Sloan Fellowship. JT acknowledges that support for this work was provided by NASA through the NASA Hubble Fellowship grant \#51424 awarded by the Space Telescope Science Institute, which is operated by the Association of Universities for Research in Astronomy, Inc., for NASA, under contract NAS5-26555. This project was developed in part at the 2017 Heidelberg Gaia Sprint, hosted by the Max-Planck-Institut f\"ur Astronomie, Heidelberg, and the 2018 NYC Gaia Sprint, hosted by the Center for Computational Astrophysics of the Flatiron Institute in New York City.

This research made use of the cross-match service provided
by CDS, Strasbourg. Analyses and plots presented in this article
used \texttt{iPython}, and packages in the \texttt{SciPy} ecosystem
\citep{Jones:2001aa,4160265,4160251,5725236}.  The study made use
of high performance computing facilities at Liverpool John Moores
University, partly funded by the Royal Society and LJMU's Faculty
of Engineering and Technology.

Funding for the Sloan Digital Sky Survey IV has been provided by the Alfred P. Sloan Foundation, the U.S. Department of Energy Office of Science, and the Participating Institutions. SDSS-IV acknowledges
support and resources from the Center for High-Performance Computing at
the University of Utah. The SDSS web site is www.sdss.org.

SDSS-IV is managed by the Astrophysical Research Consortium for the 
Participating Institutions of the SDSS Collaboration including the 
Brazilian Participation Group, the Carnegie Institution for Science, 
Carnegie Mellon University, the Chilean Participation Group, the French Participation Group, Harvard-Smithsonian Center for Astrophysics, 
Instituto de Astrof\'isica de Canarias, The Johns Hopkins University, 
Kavli Institute for the Physics and Mathematics of the Universe (IPMU) / 
University of Tokyo, the Korean Participation Group, Lawrence Berkeley National Laboratory, 
Leibniz Institut f\"ur Astrophysik Potsdam (AIP),  
Max-Planck-Institut f\"ur Astronomie (MPIA Heidelberg), 
Max-Planck-Institut f\"ur Astrophysik (MPA Garching), 
Max-Planck-Institut f\"ur Extraterrestrische Physik (MPE), 
National Astronomical Observatories of China, New Mexico State University, 
New York University, University of Notre Dame, 
Observat\'ario Nacional / MCTI, The Ohio State University, 
Pennsylvania State University, Shanghai Astronomical Observatory, 
United Kingdom Participation Group,
Universidad Nacional Aut\'onoma de M\'exico, University of Arizona, 
University of Colorado Boulder, University of Oxford, University of Portsmouth, 
University of Utah, University of Virginia, University of Washington, University of Wisconsin, 
Vanderbilt University, and Yale University.

This work has made use of data from the European Space Agency (ESA) mission Gaia (http://www.cosmos.esa.int/gaia), processed by the Gaia Data Processing and Analysis Consortium (DPAC, http://www.cosmos.esa.int/web/gaia/dpac/consortium). Funding for the DPAC has been provided by national institutions, in particular the institutions participating in the Gaia Multilateral Agreement.
%%%%%%%%%%%%%%%%%%%%%%%%%%%%%%%%%%%%%%%%%%%%%%%%%%

%%%%%%%%%%%%%%%%%%%% REFERENCES %%%%%%%%%%%%%%%%%%

% The best way to enter references is to use BibTeX:

\bibliographystyle{mnras}
\bibliography{bib} % if your bibtex file is called example.bib

% Alternatively you could enter them by hand, like this:
% This method is tedious and prone to error if you have lots of references
%\begin{thebibliography}{99}
%\bibitem[\protect\citeauthoryear{Author}{2012}]{Author2012}
%Author A.~N., 2013, Journal of Improbable Astronomy, 1, 1
%\bibitem[\protect\citeauthoryear{Others}{2013}]{Others2013}
%Others S., 2012, Journal of Interesting Stuff, 17, 198
%\end{thebibliography}

%%%%%%%%%%%%%%%%%%%%%%%%%%%%%%%%%%%%%%%%%%%%%%%%%%

%%%%%%%%%%%%%%%%% APPENDICES %%%%%%%%%%%%%%%%%%%%%

\appendix

\section{Estimating ages for Red giant stars using Bayesian COnvolutional Neural Networks}
\label{sec:appA}
The APOKASC-2 catalogue \citep{2018arXiv180409983P} consists of asteroseismic and spectroscopic data products measured for 6,676 stars in common between the APOGEE and NASA-\emph{Kepler} mission. This represents a significantly larger catalogue than the previous release, as well as a number of improvements over the original analysis \citep[presented and examined in][]{2014ApJS..215...19P}.  APOKASC DR2 derives stellar properties by combining spectroscopically measured parameters from APOGEE DR14 with light curves from \emph{Kepler}, the improved reduction of which is described by Elsworth et al. (2018, in prep.). The asteroseismic analysis is performed using a set of five distinct pipelines, where the mean results of these are used to generate the final catalogue. Pipelines estimate the asteroseismic parameters, the large frequency separation $\Delta \nu$ and the frequency of maximum power, $\nu_\mathrm{max}$, which are then used in semi-empirical scaling relations that give the mass and radius of the star, given a known $T_\mathrm{eff}$ (the spectroscopic data). Scatter about the ensemble mean of the pipelines is used to infer the random uncertainties on the asteroseismic parameters $\Delta \nu$ and $\nu_\mathrm{max}$, which are propagated through to the stellar properties. Any differences between the pipelines that depend on $\Delta \nu$ and $\nu_\mathrm{max}$ contribute to a systematic error budget. A key difference between APOKASC-1 and 2 lies in the treatment of the $\Delta \nu$ scaling relation, which is no longer considered to be an exact relation, and is corrected on a star-by-star basis, through the approach described by Serenlli et al. (2018, in prep.). This theoretical correction also relies on the APOGEE DR14 \feh{} and \afe{} values. The property of most importance to this work, the stellar ages, are finally estimated using the derived mass, surface gravity, and abundances with their uncertainties estimated through propagation of the uncertainties on those properties. We use the `recommended' age estimates and uncertainties as tabulated and available in \citet{2018arXiv180409983P}, without applying any external corrections.

\begin{figure}
\includegraphics[width=\columnwidth]{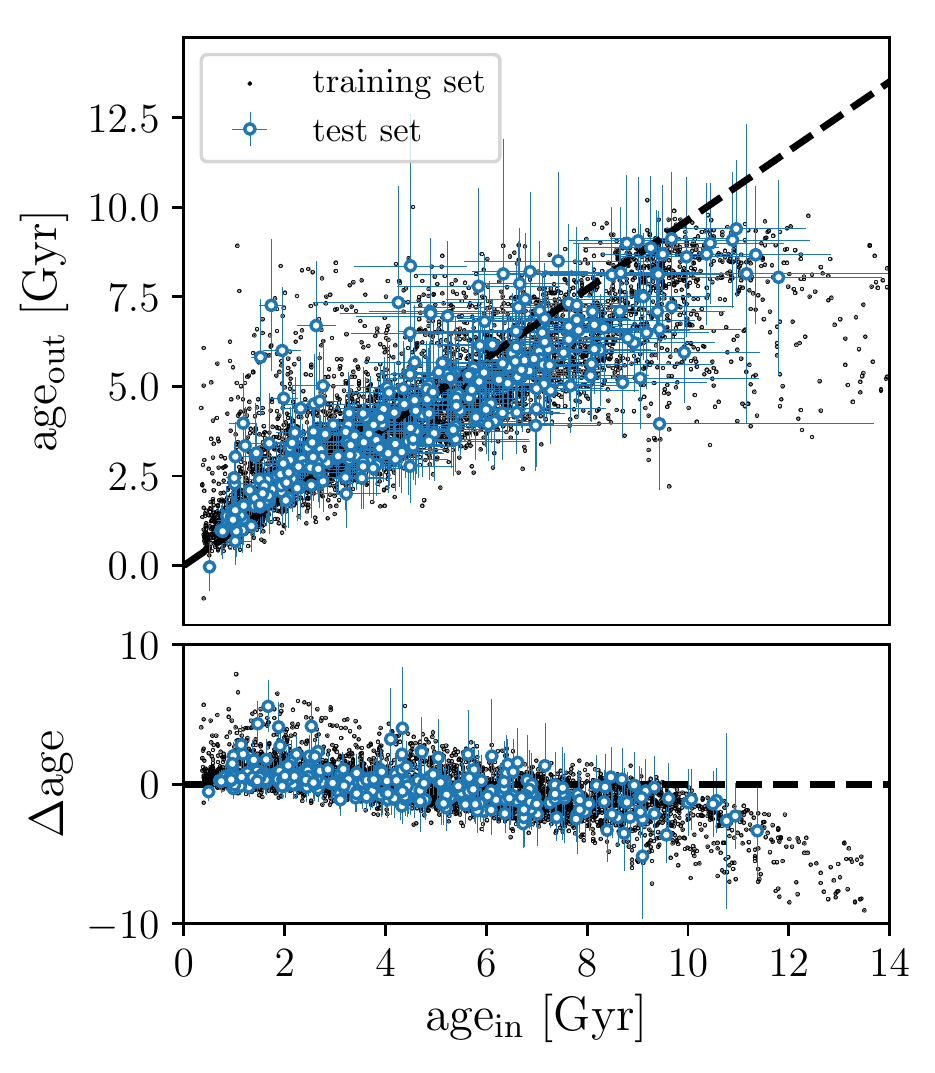}
\caption{\label{fig:ageinout}Predicted ages against the APOKASC-2 input ages for the BCNN model. The black points show the training data used, whereas the blue points and error bars reflect ages and 1$\sigma$ uncertainties for a test set, which was withheld from training at all stages. The test set generally agree with the input ages, within the uncertainties. The uncertainties tend to be larger at older ages. There is a slight trend in predicted ages, such that old ages ($\gtrsim 8$ Gyr) are slightly underpredicted by the model. The oldest ages in the training data are greatly underpredicted by the model, in a similar way as seen in other data-driven spectroscopic age inferences (e.g., \citealt{2016MNRAS.456.3655M}), likely because spectra are insensitive to stellar mass at low stellar masses.}
\end{figure}

The APOKASC-2 ages and their corresponding combined APOGEE spectra form the basis of the training set which we use here to train a model to predict ages for the remainder of the DR14 red-giant sample. This method relies mainly on the relation between the surface abundances of red giants and their main sequence mass, which is set when the stars undergo first dredge up (FDU). Stars on the main sequence undergo hydrogen fusion by the CNO and CN cycles, in which C, N and O atoms catalyse burning, and so undergo evolution in their own relative abundances. 
%The relative amounts of CNO and CN processed material are a function of the temperature of the stellar core, which increases as a function of stellar mass. Thus, when stars undergo dredge up after leaving the main sequence, the material which is brought to the surface reflects the mass of the star at the time it was produced (on the main sequence). 
 As shown by \citet{2015A&A...583A..87S}, the abundance profile of Nitrogen going inwards to the stellar core (before FDU) reflects first that of the CN cycle, followed by the CNO cycle. The FDU mixes the envelope with the CN layer. Stellar models with increased mass tend to have larger convective envelopes, meaning that they reach deeper into the stellar interior, meaning that the surface $\mathrm{[C/N]}$ decreases with increasing RGB mass.  We assume here that such differences in surface abundances of C, N and O are reflected in APOGEE spectra, such that there should be some (likely complicated) relation between the spectra and the stellar mass and, therefore, the implied age. In \citet{2016MNRAS.456.3655M}, this relationship was exploited by fitting a polynomial feature regression model to predict age from APOGEE C, N, $T_\mathrm{eff}$, $\log{g}$ and \feh{}. Here, we use the data in the full spectrum, applying a neural network based model, in an attempt to improve age estimates by capturing the more complicated aspects of the relationship between the age and the surface abundances, and avoiding the reliance of the age estimation on the individual abundance determinations from the spectra.

Our method uses a Bayesian Convolutional Neural Network (BCNN), implemented in the \texttt{astroNN} python package \citep{2018arXiv180804428L}, which wraps the \texttt{Keras} and \texttt{TensorFlow} machine learning architectures. BCNNs treat the more commonly used Convolutional Neural Network (CNN) as a Bayesian regression problem, inferring the probability distributions over the model weights. This not only allows for the propagation of uncertainty into any predictions of the model, but also offers increased robustness to overfitting, which is common in CNNs with small training data-sets (as is the case here). The use of BCNNs for the prediction of abundances for APOGEE spectra is explored in \citet{2018arXiv180804428L}. Here we focus only on their use for predicting stellar ages. 

The training data is compiled and loaded using functions in \texttt{astroNN}. Individual APOGEE visit spectra are recombined and continuum normalised using a method similar to that used in \emph{the Cannon} \citep{2015ApJ...808...16N}, which makes a Chebyshev polynomial fit to specifically selected pixels separately between the different CCD chips in APOGEE. This procedure results in an improved normalisation, which is preferable when training the BCNN model, over the normalisation used in the standard APOGEE data reduction. The model consists of 2 fully connected hidden layers, with 128 and 64 nodes (in that order), activated by the ReLU function \citep{Nair:2010:RLU:3104322.3104425}. Before this, two convolution layers process the spectra through 2, then 4 filters. Training data is fed to the model in batches of 32 spectra, and the model is trained for 60 epochs (where one epoch represents 166 batches in total, for a training set size of 5311 spectra, with a set of 590 validation spectra). The model is trained using a mean-squared-error loss function, with an initial learning rate of 0.005, which is reduced when the loss reaches a plateau. 500 spectra are retained from the training altogether, and used as a test set.

Figure \ref{fig:ageinout} shows the BCNN predicted age, $\mathrm{age}_\mathrm{out}$, against the input age from APOKASC-2, $\mathrm{age}_\mathrm{in}$, for the training (black points) and test samples (shown in blue, with their associated uncertainties, as estimated through the Bayesian dropout variational inference). This data is included as supplementary material. The bottom panel gives the residual $\Delta \mathrm{age}$ defined as $\mathrm{age}_\mathrm{out}-\mathrm{age}_\mathrm{in}$. The standard deviation of residuals across the whole range is 36\%, corresponding to a median $\Delta \mathrm{age} = 1.26$ Gyr. For input ages less than 5 Gyr, the scatter is reduced to 30\%. It is clear, however that there is a signifcant trend such that young ages are over-estimated, and old ages are under-estimated, with a stronger effect at old age. Stars with input ages of $\sim10$ Gyr are underestimated in the test set by as much as 3.5 Gyr, with a large scatter, such that no stars are correctly predicted or over-predicted. This behaviour has been seen in other methods that exploit asteroseismology, using simpler polynomial regression method based on abundance measurements to measure ages \citep[e.g.][]{2016MNRAS.456.3655M}. It is noteworthy that it persists even when the full scope of information in the spectra are used. We also tested the effect on the predicted ages of adjustments to the training data, implying progressively larger cuts in the asteroseismic $\nu_{\mathrm{max}}$ in order to remove stars which lie high on the RGB, where the asteroseismic models are less well tested. We find that implying these cuts makes little difference to the final results of the training, and the output ages from the test set match well with those predicted from the full training set in all cases. Dividing the sample between RC and RGB stars (as determined by the asteroseismology) and training the BCNN separately on these populations also provides comparable results, implying that the neural network is able to capture this information from the spectra and account for it when training on the entire APOKASC-2 sample \citep[the presence of such information in stellar spectra has been discussed by][]{2018ApJ...853...20H,2018ApJ...858L...7T}.

\begin{figure}
\includegraphics[width=\columnwidth]{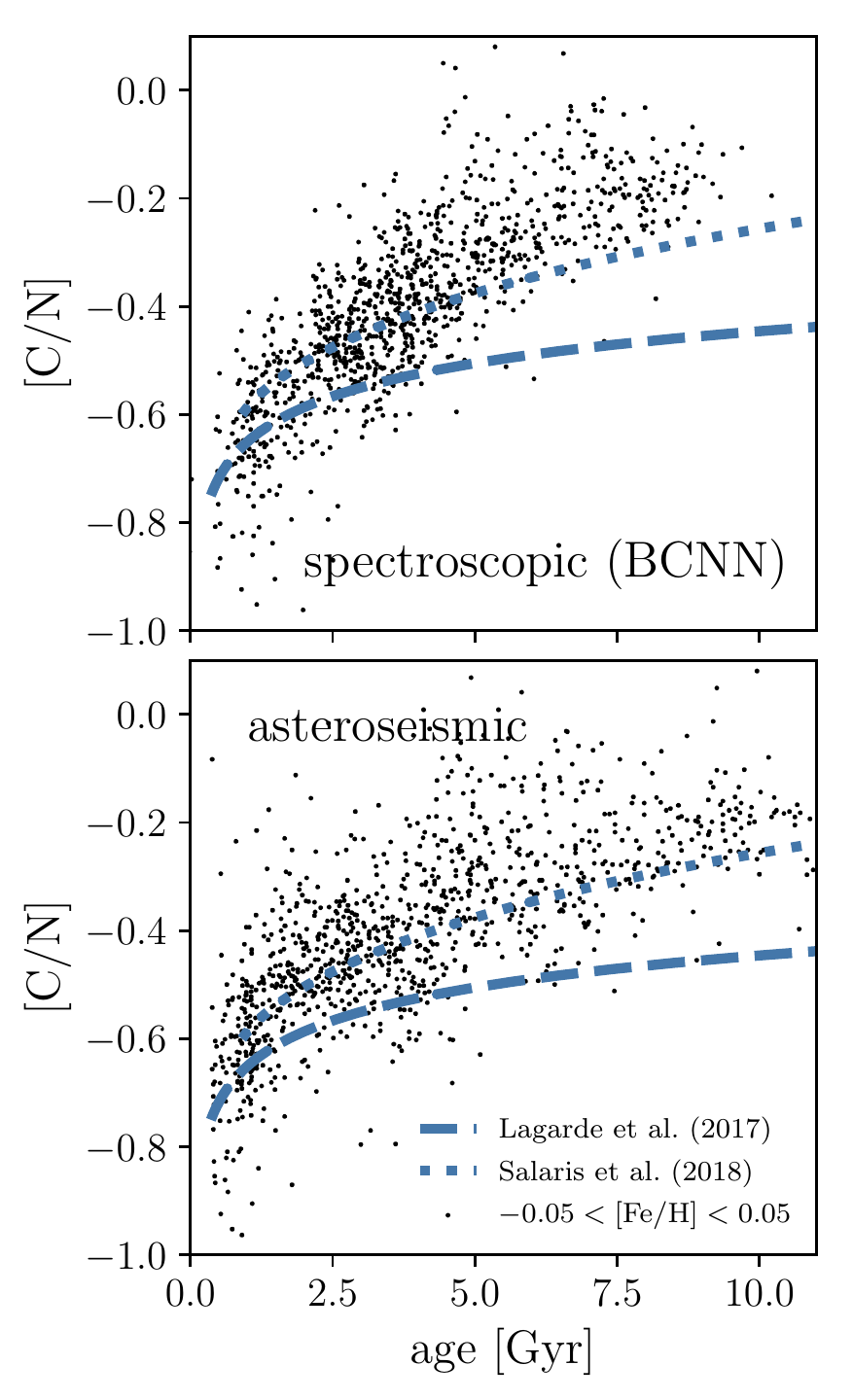}
\caption{\label{fig:cnrelations} Stellar ages predicted by the BCNN (top panel) and taken from APOKASC-2 (lower panel) plotted against the APOGEE DR14 $\mathrm{[C/N]}$ for the stars in the APOKASC-2 catalogue with $-0.05 < \mathrm{[Fe/H]} < 0.05$ dex. We overplot the age-$\mathrm{[C/N]}$ relationships as obtained for solar metallicity stars from the BASTI \citep{2015A&A...583A..87S} and STAREVOL \citep{2017A&A...601A..27L} stellar evolution models. Both models roughly match the data at young ages ($\lesssim 5$ Gyr) but begin diverging at older ages. The \citet{2017A&A...601A..27L} models appear to have a systematically low $\mathrm{[C/N]}$ at any given age. The \citet{2015A&A...583A..87S} agree well with the asteroseismic ages (the training data), whereas the spectroscopic age predictions appear to have underpredicted ages given their $\mathrm{[C/N]}$.}
\end{figure}

We further examine the issue with the absolute scaling of ages predicted from the neural network (and other) models by comparing the predicted age-$\mathrm{[C/N]}$ relationship with that obtained from stellar evolution models. In Figure \ref{fig:cnrelations}, we show the BCNN predicted ages (upper panel) and the training set APOKASC-2 ages (lower panel) against the APOGEE DR14 $\mathrm{[C/N]}$ for stars in APOKASC-2 with  $-0.05 < \mathrm{[Fe/H]} < 0.05$ dex. The age-$\mathrm{[C/N]}$ relationship for solar metallicity stars, as obtained from the BASTI \citep{2015A&A...583A..87S} and STAREVOL \citep{2017A&A...601A..27L} models, is overplotted in the dotted and dashed lines. It is apparent that the models roughly agree with the training and predicted data at young ages ($\lesssim 5$ Gyr). The \citet{2015A&A...583A..87S} model appears to agree slightly better than that of \citet{2017A&A...601A..27L}, which has a systematic offset to lower $\mathrm{[C/N]}$ at any given age. At older ages, the spectroscopic ages have higher $\mathrm{[C/N]}$ than is predicted by the models, whereas the asteroseismic input ages seem to agree well over the full range of abundance ratios. This test provides a further indication that the oldest ages are under-predicted by the neural netwok model, as the $\mathrm{[C/N]}$ ratios of these stars (as measured by APOGEE) should correspond an older age, given the stellar evolution models. This implies one of two possible explanations, either: a) the ages of old stars provided in the training set are underestimated, or b) that the spectroscopic information is not representative of the true mass (as measured by asteroseismology) at old age. As the training set appear to match the models relatively well, it would seem that the second explanation is more likely. Fully understanding this effect is beyond the scope of this paper, but will be essential for accurately determining the ages of the oldest stars in the Milky Way using spectroscopy.

 In a previous paper in this series, \citet{2017arXiv170600018M}, we chose to correct for the under-prediction of old ages using a non-parametric LOWESS fit to the $\mathrm{age}_\mathrm{out}-\mathrm{age}_\mathrm{in}$ relation. Here, we leave under-predicted ages in the full data set, and simply warn readers that results pertaining to any trends in age at values greater than $\sim 8$ Gyr likely extend from that age to the oldest stars in the sample. The above tests demonstrate that the sample likely contains stars older than the $\sim 10$ Gyr limit seen in the final predicted ages, above which it appears the spectroscopic information is no longer useful to predict ages. The final age catalogue for DR14 stars, as predicted by the BCNN, is included in a supplementary data table.

\section{Example posterior PDF of velocity distribution model}
\label{sec:appB}
 {
To demonstrate the overall quality of the fit to the velocity distribution that can be achieved using the APOGEE-\emph{Gaia} sample, we provide here an example of the posterior probability distribution gained from the MCMC sampling of the parameters for the velocity distribution of a single exemplary bin in age-\feh{}. The bin which we choose to show here is at intermediate age ($4.0 < \mathrm{age} < 4.5\ \mathrm{Gyr}$ old) and sub-solar \feh{} ($-0.1 < \mathrm{[Fe/H]} < 0.$) in the low \afe{} population, which has 2178 stars. The posterior distribution of each of the parameters and their covariance is shown in a corner plot, in Figure \ref{fig:posterior}. The posteriors are in most cases well behaved and approximately Gaussian. As expressed in Section \ref{sec:models}, decreasing the number of stars used for the fit broadens the distributions, but the median values remain the same within the uncertainties. Checking the posterior distributions for all the mono-age, mono-\feh{} bins 

\begin{figure*}
\includegraphics[width=\textwidth]{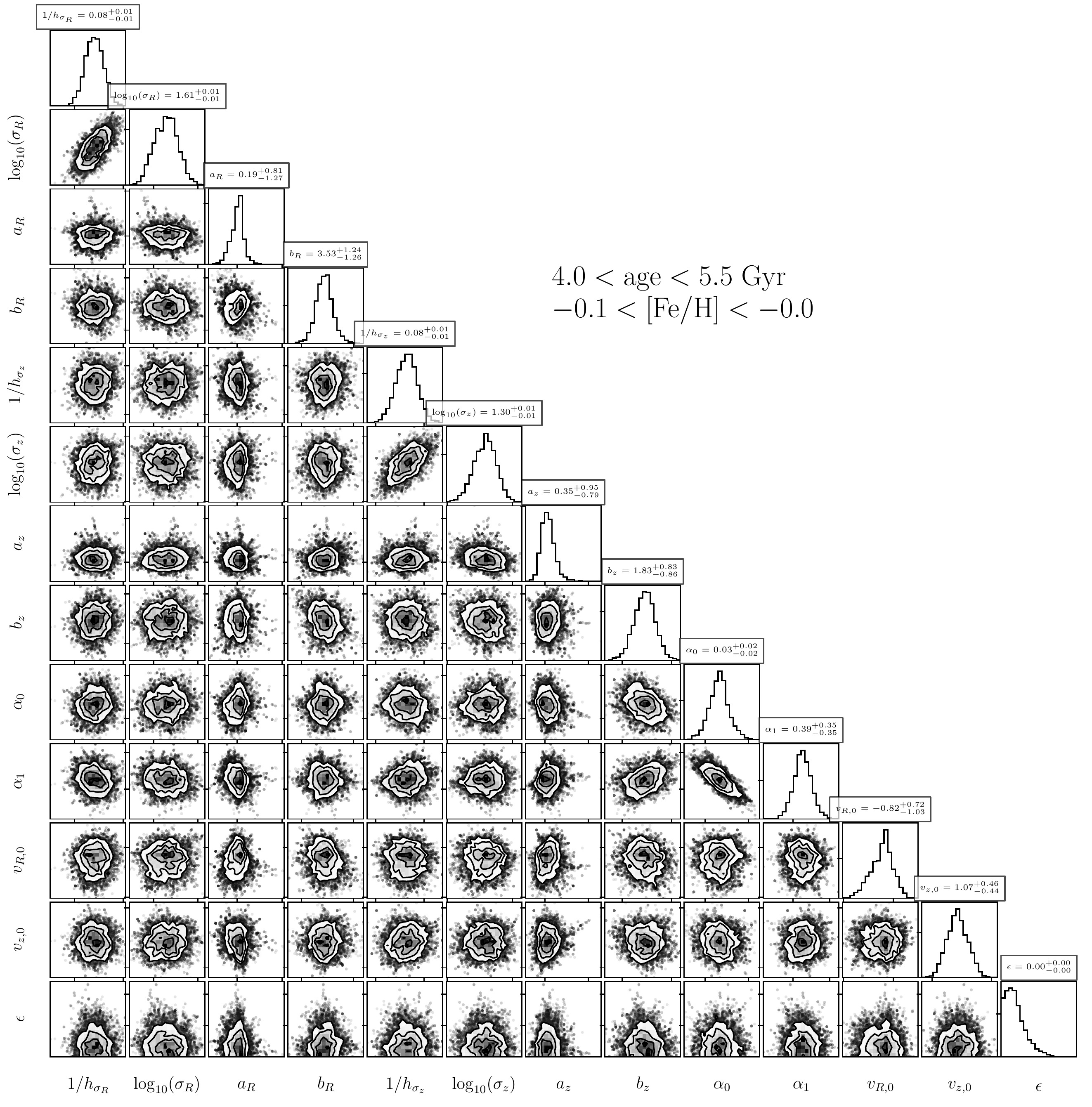}
\caption{\label{fig:posterior} Posterior distributions of the parameters for a single, exemplary, mono-age, mono-\feh{} bin in the low \afe{} population with $4.0 < \mathrm{age} < 4.5\ \mathrm{Gyr}$ old, and $-0.1 < \mathrm{[Fe/H]} < 0.$. The posteriors are generally well behaved and approximately Gaussian, with little covariance between most of the parameters.}
\end{figure*}

}
%%%%%%%%%%%%%%%%%%%%%%%%%%%%%%%%%%%%%%%%%%%%%%%%%%

% Don't change these lines
\bsp	% typesetting comment
\label{lastpage}
\end{document}